\documentstyle[epsf]{l-aa}

\begin{document}
  \thesaurus{12.          % A&A Section 12: Physical processes
              (08.14.1;   % Stars: neutron
               08.16.6;   % pulsars: general
               02.04.1)   % Dense matter
            }
\title{Internal temperatures and cooling 
of neutron stars \protect\\
with accreted envelopes}

\author{A.Y.\,Potekhin$^{1,2}$\thanks{
E-mail: palex@astro.ioffe.rssi.ru} 
\and G.\,Chabrier$^2$ 
\and D.G.\,Yakovlev$^1$
}
\institute{
$^1$ Ioffe Physical-Technical Institute, 
     Politekhnicheskaya 26,
     194021 St.-Petersburg, Russia\\
$^2$ Centre de Recherche Astronomique de Lyon (UMR CNRS \# 5574), 
     Ecole Normale Sup\'erieure de Lyon,
     69364 Lyon Cedex 07, France
}
\offprints{A.Y.\,Potekhin (Ioffe Institute)}

\date{Received 2 October 1996 / Accepted 23 December 1996}

\maketitle

\markboth{A.Y.\,Potekhin, G.\,Chabrier, D.G.\,Yakovlev: 
Cooling of neutron stars with accreted envelopes}{}

\begin{abstract}
The relationships between the effective surface
($T_{\rm eff}$) and internal
temperatures of neutron stars (NSs)
with and without accreted envelopes
are calculated for $T_{\rm eff} > 5\times10^4$~K
using new data on
the equation of state and opacities in the outer NS layers.
We examine various models of accreted layers
(H, He, C, O shells produced by nuclear transformations
in accreted matter). We employ new Opacity Library
(OPAL) radiative opacities for H,
He, and Fe. In the outermost NS layers,
we implement the modern OPAL equation of state for Fe,
and the Saumon--Chabrier equation of state for H and He.
The updated thermal conductivities of degenerate electrons include
the Debye--Waller factor for the electron-phonon scattering
in solidified matter, while in liquid matter
they include the contributions
from electron-ion collisions
(evaluated with non-Born corrections and with the ion structure factors
in responsive electron background)
and from the electron-electron collisions.
For $T_{\rm eff} < 10^{5.5}$ K,
the electron conduction in non-degenerate layers of the envelope
becomes important, reducing noticeably the temperature gradient.
The accreted matter further decreases
this gradient at $T_{\rm eff} > 10^5$ K.
Even a small amount of accreted matter
(with mass $\ga 10^{-16} M_\odot$) affects appreciably
the NS cooling, leading
to higher $T_{\rm eff}$ at the neutrino cooling stage and
to lower $T_{\rm eff}$ at the subsequent photon stage.


\end{abstract}

%%  SECTION 1 *********************************************************
\section{Introduction}
% ---------------------------------------------------------------------

It is well known that the cooling of a neutron star (NS) is
strongly affected by the relationship between the
internal temperature of the star, $T_{\rm b}$,
and its effective surface temperature $T_{\rm eff}$.
Throughout this paper, $T_{\rm b}$ denotes the temperature
at the outer boundary of the isothermal internal region.
The $T_{\rm b}$--$T_{\rm eff}$ relationship is
determined by equation of state (EOS) and thermal conductivity of
matter in the outer NS envelope.
For non-magnetized NSs with envelopes composed of iron,
this relationship has been thoroughly studied
in a classical article of Gudmundsson
et al.\ (1983, hereafter GPE). Several papers
(Hernquist 1984, 1985, Van Riper 1988, 1991, Schaaf 1988, 1990)
have considered strongly magnetized NS envelopes.

In this article, we will reconsider thermal transport
in non-magnetized envelopes of NSs and extend the results
of the preceding authors in two respects.

First, we analyze matter composed
not only of iron, but of light elements as well.
The light elements can be provided by an accretion
from a supernova remnant
(e.g., Chevalier 1989, 1996, Brown \& Weingartner 1994),
from interstellar medium (e.g.,
Miralda-Escud\'{e} et al.\ 1990, Blaes et al.\ 1992, 
Nelson et al.\ 1993, Morley 1996),
from a distant binary component, or by comets.
Freshly accreted matter burns then
into heavier elements (He, C, O, Fe)
while sinking within the NS. Recent multiwavelength observations
of the Geminga pulsar (1E 0630+17.8)
suggest a possible H or He cyclotron
feature in its spectrum (Bignami et al.\ 1996).
If confirmed, it may be a direct observational
evidence of the presence of the light elements
in the pulsar atmosphere.
Chemical composition affects the EOS and thermal conduction,
and, therefore, the thermal structure and cooling of a NS.
As a reference case, we reconsider the outer NS envelopes
composed of iron.

Second, we implement new, advanced
theoretical data on EOS and thermal
conductivity of dense matter. Specifically,
we employ the Opacity Library (OPAL) radiative opacities for H,
He, and C, improved considerably
with respect to 
the Los-Alamos opacities used in the previous studies.
In the outermost NS layers, we also implement
the modern OPAL EOS for Fe
and the Saumon--Chabrier EOS for H and He.
We use improved thermal conductivities of degenerate
electrons. For solidified matter, we employ the thermal conductivity
due to the electron-phonon scattering obtained with the inclusion
of the Debye--Waller factor. For liquid matter, we recalculate
and implement the thermal conductivity due to Coulomb electron-ion
and electron-electron collisions.
The electron-ion scattering is described with the exact (non-Born)
Coulomb cross sections
and with the ion structure factors
calculated when taking into account the response of the electron background.
The new physics input allows us to extend the results of previous
studies to colder NSs, with $T_{\rm eff}$ down to 50\,000~K.

The physics input is described in Sec.\ 2.
In Sec.\ 3, we calculate the $T_{\rm b}$--$T_{\rm eff}$ relationships
for non-accreted and partly accreted NSs and analyze the
sensitivity of the results to the uncertainty in our knowledge of
the electron thermal conductivity. In addition, we simulate
the cooling of NSs with standard and enhanced neutrino
energy losses. We show that the cooling of a
NS with the accreted envelope can be quite different
from the cooling of a non-accreted NS.
This can change the conclusions on the internal structure
of NSs deduced from comparison of theoretical
cooling curves with observations of NS thermal radiation.
Useful analytical formulae for the physics input are given in the 
Appendix.

%  Section 2 ********************************************************
\section{Physical input}

%  Sec. 2.1 ---------------------------------------------------------
\subsection{Thermal structure equation}
\label{sect-th-str-eq}
% -------------------------------------------------------------------

Consider thermal transport throughout the outer envelope
of a NS that extends from the surface 
to the layers with the density
$\rho_{\rm b} \la 10^{10}$ g cm$^{-3}$.
This envelope, of which study is the aim
of the present paper,
provides the main thermal insulation of the NS interior.
The envelope is thin ($\sim 10^2$ m deep) and contains
very small fraction ($\sim 10^{-7}$) of the NS mass (GPE).
In these layers, one can neglect the nonuniformity
of the energy flux due to the neutrino emission
and the variation of the gravitational acceleration.
Then the temperature profile
obeys the {\it thermal structure equation\/} (GPE),
which can be written as (e.g., Van Riper 1988)
\begin{equation}
   {{\rm d}\log T\over{\rm d}\log P} =
   {3\over 16}\,{PK\over g}\,{T_{\rm eff}^4\over T^4}.
\label{th-str}
\end{equation}
Here, $P$ is the pressure, $T_{\rm eff}$ is the effective temperature,
$g=GM/(R \, \sqrt{1-r_{\rm g}/R})$ is the surface gravity,
$M$ is the stellar mass, $R$ the stellar radius,
$G$ the gravitational constant,
and $r_{\rm g} = 2GM/c^2$ is the gravitational radius of the star.
Furthermore, $K$ is the opacity,
\begin{equation}
   K=\left( {1 \over K_{\rm rad}}+
            {1 \over K_{\rm c}} \right)^{-1},
\label{K}
\end{equation}
composed of the radiative opacity $K_{\rm rad}$
and the equivalent electron conduction opacity $K_{\rm c}$.
The latter is related to the electron thermal conductivity $\kappa$ as
\begin{equation}
   K_{\rm c}={16\sigma T^3 \over 3\rho\kappa},
\label{K_c}
\end{equation}
where $\sigma$ is the Stefan--Boltzmann constant.

The effective temperature $T_{\rm eff}$
is defined through the Stefan's law,
\begin{equation}
   {\cal L}_{\rm rad} = 4\pi R^2\sigma T_{\rm eff}^4,
\end{equation}
where ${\cal L}_{\rm rad}$
is the local radiative luminosity. The apparent luminosity
measured by a distant observer is
${\cal L}_\infty=(1-r_{\rm g}/R) \, {\cal L}_{\rm rad}$,
and the apparent surface temperature inferred
by the observer from the spectrum is
$T_\infty=T_{\rm eff} \sqrt{1-r_{\rm g}/R}$ (e.g., Thorne 1977).

We adopt the standard (GPE)
outer boundary condition to Eq.~(\ref{th-str})
by equating $T_{\rm eff}$ to the temperature
$T_{\rm s}$ at the stellar surface
in the Eddington approximation. The radiative
surface, determined in this way, lies at the optical
depth $\tau=2/3$, that is approximately at 
\begin{equation}
   P_{\rm s}\approx (2/3)\,g/K_{\rm s}.
\label{Ps}
\end{equation}
Here and hereafter the subscript ``s'' denotes quantities
taken at the surface, while the subscript ``b'' is assigned
to those at the inner boundary $\rho=\rho_{\rm b}$. 
Previous calculations of the radiative transfer in the outer part 
of the atmosphere (Zavlin et al.\ 1996) have shown that, 
for the parameters of interest, there 
is no convective instability at $\tau < 1$ which could invalidate 
the Eddington approximation.  

Starting from the surface specified by Eq.~(\ref{Ps}), we integrate
Eq.\,(\ref{th-str}) inward, using the classical 4-step 4th-order
Runge--Kutta algorithm (e.g., Fletcher 1988). The integration step
varies with depth and temperature to ensure the variations of
$T$ and $K$ to be smaller than 5\% at one step.
Following previous studies
(e.g., GPE, Hernquist 1985, Van Riper 1988), we terminate the integration at
$\rho_{\rm b}=10^{10}$ g cm$^{-3}$. 
A comparison with analytically solvable models
(e.g., Hernquist \& Applegate 1984) shows that
our numerical procedure determines $T_{\rm b}$ at a given
$T_{\rm s}$ with an error $\la$ 1\%.

% Sec. 2.2  %%%%%%%%%%%%%%%%%%%%%%%%%%%%%%%%%%%%%%%% SUBSECTION
\subsection{Equation of state}
% -------------------------------------------------------------

% Sec. 2.2.1 %%%%%%%%%%%%%%%%%%%%%%%%%%%%%%%%%%%%%%% SUBSUBSECTION
\subsubsection{Plasma parameters}
% --------------------------------------------------------------

Although the density $\rho$ does not enter Eq.\,(\ref{th-str})
explicitly, its value is required to determine the opacity $K$
at given $(T,P)$. Thus, we need
an EOS of matter in the outer NS envelope.
In this paper, we are specifically interested
in H, He, C, O, and Fe plasmas,
which are likely to compose partly accreted NS envelopes.
The parameters of interest cover the region which extends
from the partially ionized non-degenerate non-relativistic
plasma near the surface, through the domain of pressure ionization,
to high-density layers of fully ionized relativistic plasma.
We assume a strongly stratified envelope, i.e., 
that at each given density, the plasma is composed of
electrons and of ions of
one chemical element which can be in different ionization stages.

The plasma density can be characterized by the parameter
$
   r_s=a_{\rm e}/a_{\rm B},
$
where $a_{\rm e}=(4 \pi n_{\rm e}/3)^{-1/3}$
is the mean inter-electron distance,
$n_{\rm e}$ is the number density of electrons,
and $a_{\rm B}$ is the Bohr radius.
The parameter $r_s$ is directly related
to the relativistic parameter of degenerate electrons
$x\equiv p_{\rm F}/m_{\rm e}c
=1.009(\rho_6 \langle Z\rangle/\langle A\rangle)^{1/3}$
(e.g., Yakovlev \& Shalybkov 1989):
$
   r_s=0.0140 / x
$.
Here, $p_{\rm F}=\hbar(3\pi^2n_{\rm e})^{1/3}$
is the electron Fermi momentum, $m_{\rm e}$ is the electron mass,
$\rho_6 \equiv \rho/10^6$ g cm$^{-3}$, and
$\langle Z \rangle$ and $\langle A\rangle$ are
the mean charge and mass
number of ions, respectively. In our case, the averaging is
over different ionization stages of ions of the same chemical
element.

The electron degeneracy temperature is given by
\begin{equation}
   T_{\rm F}=\left[
   \sqrt{1+x^2}-1\right] m_{\rm e}c^2
   / k_{\rm B},
\end{equation}
where $k_{\rm B}$ is the Boltzmann constant.

Let us also introduce the ion coupling parameter,
   $\Gamma= \langle Z^{5/3}\rangle e^2/ (a_{\rm e} k_{\rm B} T)$,
which measures typical energy
of ion Coulomb interaction relative to
the thermal energy ($e$ being the electron charge).

% Sec. 2.2.2 %%%%%%%%%%%%%%%%%%%%%%%%%%%%%%%%%%%%%%%% SUBSUBSECTION
\subsubsection{High density regime}
\label{sect-highdens}
%-----------------------------------------------------------------

Let us specify the high density regime as $r_s< 1/(2Z)$.
This corresponds to
$\rho > \rho_{\rm h} = 21.6 \, Z^2 A$ g cm$^{-3}$,
$Z$ and $A$ being the charge and mass numbers of atomic nuclei.
In this regime, $a_{\rm e}$ is considerably smaller than
the $K$-shell radius of an atom, and we have 
a fully ionized one-component plasma (OCP) of ions immersed in the
``rigid'' electron background. Then the pressure is
\begin{equation}
   P=P_{\rm e}+P_{\rm i}+P_{\rm C},
\label{Phigh}
\end{equation}
where $P_{\rm e}$ is the electron pressure, $P_{\rm i}$
is the pressure of ideal gas of ions,
and $P_{\rm C}$ is the Coulomb term.

The pressure of the ideal gas of electrons of any
degeneracy is given by
(e.g., Landau \& Lifshitz 1986),
\begin{equation}
   P_{\rm e}={k_{\rm B} T\over\pi^2\hbar^3}\int_0^\infty
   \ln\left[1+\exp\left({\mu-\epsilon\over k_{\rm B} T}\right)\right]
   p^2{\rm d}p,
\label{P_e}
\end{equation}
where $\epsilon= \sqrt{m_{\rm e}^2 c^4 +p^2c^2}$
is the energy of an electron with a momentum $p$.
The electron chemical potential $\mu$ has to be determined
from the equation
\begin{equation}
   n_{\rm e} = {1\over\pi^2 \hbar^3}
   \int_0^\infty { p^2 {\rm d} p \over
   1+\exp[(\epsilon-\mu)/(k_{\rm B}T)]}.
\label{ne}
\end{equation}

The pressure contribution from the
ideal gas of ions in Eq.\,(\ref{Phigh}) is
\begin{equation}
   P_{\rm i}=n_{\rm i}k_{\rm B}T,
\label{P_i}
\end{equation}
where $n_{\rm i}$ is the ion number density.

The Coulomb correction accounts for electrostatic
interactions of plasma particles.
According to Hansen \& Vieillefosse (1975),
the results of Monte Carlo simulations of the OCP of ions
immersed in a rigid electron background can be fitted by
\begin{equation}
   P_{\rm C}=P_{\rm i}\,\Gamma^{3/2}
   \left[{A_1\over (B_1+\Gamma)^{1/2}}
   + {A_2\over B_2+\Gamma}\right],
\label{P_C}
\end{equation}
where $A_1=-0.899962$, $B_1=0.702482$, $A_2=0.274105$,
and $B_2=1.319505$.
The fit error is smaller than 1\% at $\Gamma>1$,
as confirmed by comparison with the recent Monte Carlo results
of Stringfellow et al.\ (1990) and DeWitt et al.\ (1996).
Equation (\ref{P_C}) reproduces
also the correct Debye--H\"uckel limit at $\Gamma \ll 1$.

Equations (\ref{Phigh})--(\ref{P_C}) cannot be used
at $r_s \ga Z^{-1}$ and $\Gamma \ga 1$ because of the failure
of the rigid electron background approximation.
They lead even to negative pressure
at low temperatures and intermediate densities.
Van Riper (1988) modified the Coulomb correction (\ref{P_C})
in an {\it ad hoc\/} manner
to avoid negative pressures. We use another approach
described below in Sect.~\ref{sect-interdens}.

% Sec. 2.2.3 %%%%%%%%%%%%%%%%%%%%%%%%%%%%%%%%%%% SUBSUBSECTION
\subsubsection{Low density regime}
\label{sect-lowdens}
In the {\it low density regime\/}, $r_s \gg 1$ and $\Gamma \ll 1$,
the plasma is nearly ideal and may be partially ionized,
depending on $T$ and $\rho$.
Theoretical description of partially ionized plasmas
can be based either on the physical picture or
on the chemical picture  
of the plasma (e.g., Ebeling et al.\ 1977).
In the chemical picture, the bound species (atoms, molecules, ions) are
treated as elementary members of the thermodynamic ensemble, along
with free electrons and nuclei. In the physical picture, nuclei and
electrons (free and bound) are the only constituents of the
ensemble. Both pictures can be thermodynamically self-consistent,
though the chemical picture has limited microscopic consistency
(see, e.g., Chabrier \& Schatzman 1994, for review).

The most advanced results based on the physical picture
are those obtained in the OPAL project (Rogers et al.\ 1996).
However, as an intrinsic limitation of the formalism,
they are restricted to low densities and/or high temperatures.
We employ the OPAL EOS for partially ionized iron,
which composes non-accreted envelopes.

For the outermost layers of accreted envelopes, composed of H
and He, we use the EOS derived by Saumon et al. (1995) (SCVH)
within the framework of the chemical picture.
The second-order thermodynamic quantities provided by SCVH
suffer from some thermodynamic inconsistency in some 
regions of the phase diagram (see SCVH and Potekhin 1996, 
for a discussion),
but the inconsistency is quite small at low densities.
A comparison of the SCVH and OPAL EOSs for H and He
reveals no discrepancies which could noticeably affect the
temperature profiles in a NS envelope under the conditions
of interest. Furthermore,
the SCVH tables cover a larger $\rho-T$ domain and
provide also, along with the first- and second-order
thermodynamic quantities,
the fractions of H and He atoms, H$_2$ molecules,
and H$^+$, He$^{+}$, He$^{++}$ ions,
which enables us to determine an average ion charge,
required for our calculations
of the conductivities (Sect.~\ref{sect-conduct}).

% Sec. 2.2.4 %%%%%%%%%%%%%%%%%%%%%%%%%%%%%%%%%%%% SUBSUBSECTION
\subsubsection{Intermediate density regime}
\label{sect-interdens}
%--------------------------------------------------------------

The intermediate density domain ($Z^{-1}\la r_s\la 1$)
is most complicated because of the onset of recombination.
Previous studies of the thermal structure of the NS envelopes
(GPE, Van Riper 1988) employed the model
described by Eqs.\,(\ref{Phigh})--(\ref{P_C}),
with the free electron density $n_{\rm e}$ taken from
the Los Alamos Astrophysical Opacity Library
(Huebner et al.\ 1977, hereafter LAO).
However, this approach fails at intermediate densities
and relatively low temperatures (see, e.g., Van Riper 1988),
where it leads to unrealistic EOS
owing to too large values of the Coulomb correction.
% ***
In the afore-mentioned
density region, the less dense and more strongly correlated
electron gas is polarized by the external ionic
field and eventually the electrons
become bound by pressure recombination. Equation
(\ref{P_C}) becomes inadequate in this case,
since it applies only to a {\it fully ionized} plasma, immersed
in a {\it rigid}
electron background.
% ***

Fortunately, at temperatures of present interest
($\log T\, \mbox{[K]} \ga 4.7$), the SCVH tables extend up
to $\rho>\rho_{\rm h}$ ($\sim 400$ g cm$^{-3}$ for He),
i.e., to the completely ionized region.
Thus, they fully cover the intermediate density range
for H and He. We use these tables
up to the highest tabulated densities, and we
use the high-density EOS given
by Eqs.\,(\ref{Phigh})--(\ref{P_C}) for still higher $\rho$.

The high-density domain for C and O
starts from $\rho_{\rm h}\approx 10^4$ g cm$^{-3}$.
We do not need to consider C and O
at lower densities in the present work.

The case of iron is more difficult.
The OPAL tables are bound from low-$T$ high-$\rho$ side
approximately by the line 
$\rho=\rho_{\rm OPAL}\equiv 10\, (T/10^6\,\mbox{K})^3$ g cm$^{-3}$.
In order to extend the EOS beyond this boundary,
we adopt the approach of Fontaine et al.\ (1977),
who circumvented a similar difficulty
by interpolating pressure logarithm along isotherms
over the intermediate density region.
However, the conventional cubic spline
interpolation of $\log P$ vs \ $\log\rho$
violates the condition
$
   (\partial P/\partial T)_\rho > 0.
$
Although this ``normality condition'' 
is not required by the thermodynamic consistency,
it is supposed to hold at intermediate densities 
(see, e.g., Fontaine et al.\ 1977). 

In order to keep this condition,
we introduce the following modification
of the interpolation procedure.
(i) Pressure isotherms
    from $\log T = 4.5$ to 8
    are taken from the OPAL data at $\rho < \rho_{\rm OPAL}$ 
    and calculated from
    Eqs.\,(\ref{Phigh})--(\ref{P_C})
    at $\rho > \rho_{\rm h}$.
(ii) The differences $\Delta\log P$ between $\log P$
    for neighbouring isotherms are interpolated
    across the gaps in $\log\rho$ between $\rho_{\rm OPAL}$ 
    and $\rho_{\rm h}$ by cubic splines.
(iii) Starting with the isotherm
    $\log T\mbox{(K)} = 8.25$, for which
    Eqs.~(\ref{Phigh})--(\ref{P_C}) are sufficiently accurate,
    the pressure is consecutively calculated
    between $\rho_{\rm OPAL}$ 
    and $\rho_{\rm h}$ 
    along the isotherms $\log T\mbox{(K)} = 8.0, 7.75,\ldots 4.5$,
    using the differences $\Delta\log P$ obtained at the
    preceding step: $\log P = \log P' - \Delta\log P$,
    where $P'$ is the pressure at the preceding (hotter) isotherm. 

In order to verify this procedure, we have applied it to He 
and found that 
it gives much better agreement with the SCVH EOS 
than the straightforward 
$\log P$--$\log\rho$ interpolation across the intermediate 
densities. 

% Sec. 2.2.5 %%%%%%%%%%%%%%%%%%%%%%%%%%%%%%%%%%%%%%%% SUBSUBSECTION
\subsubsection{Effective charge}
\label{sect-zeff}
% -----------------------------------------------------------------

We will treat the electron heat conduction in the mean
ion approximation, that is, we consider the plasma as
composed of electrons and one ionic species with
an effective charge $Z_{\rm eff}$.
In the case of H and He,
we adopt the mean ion charge $Z_{\rm eff}=\langle Z\rangle$
provided by the SCVH tables, as discussed above.
For carbon and oxygen, only the high-density, fully
ionized regime is involved,
for which $Z_{\rm eff}=Z=6$ and 8, respectively. Finally, for iron we
adjust the effective number density of free electrons
$n_{\rm e}$ for the pressure
$P_{\rm e}+P_{\rm i}$, calculated formally from
Eqs.~(\ref{P_e})--(\ref{P_i}), to reproduce the
pressure $P$ given by the accurate EOS at the same $\rho$ and $T$.
The mean ion charge, that corresponds to $n_{\rm e}$,
is then assumed to be equal to $Z_{\rm eff}$.

%% Fig. 1 %%%%%%%%%%%%%%%%%%%%%%%%%%%%%%%%%%%%% FIGURE
\begin{figure}[t]
\begin{center}
\leavevmode
\epsfysize=55mm
\epsfbox[70 240 540 575]{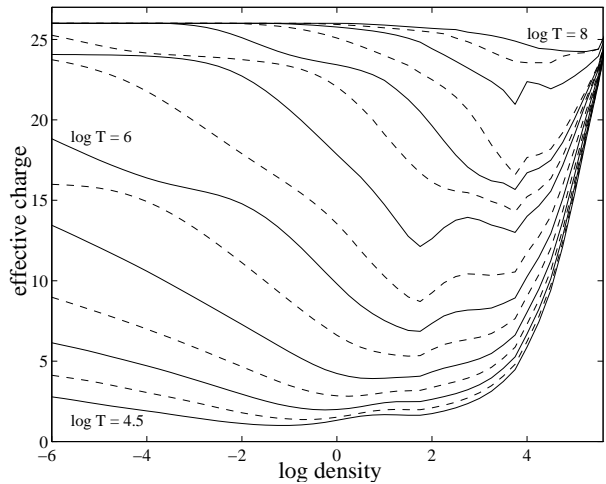}
\end{center}
\caption[]{
Isotherms of effective charge
for Fe obtained from the interpolated EOS
at log$T$[K] = 4.5 (solid line), 4.75 (dashes), 5 (solid), \dots 8.
}
\label{fig-zeff}
\end{figure}

Figure~\ref{fig-zeff} shows the isotherms of $Z_{\rm eff}$
corresponding to the interpolated Fe EOS.
The decreasing parts of the curves reflect the electron
recombination due to the reduction of phase
space per particle with increasing density.
In this low-density region, our $Z_{\rm eff}$ coincides
with $\langle Z \rangle$ given by the OPAL data (when
available) within $\sim 10$\%.
The increase of $Z_{\rm eff}$ at high densities
corresponds to pressure ionization.
The non-monotonic behaviour at intermediate densities
reflects consecutive pressure destruction
of different electron shells of Fe atoms.

%%%Sec. 2.3     %%%%%%%%%%%%%%%%%%%%%%%%%%%%%%%%% SUBSECTION
\subsection{Opacities}
%-----------------------------------------------------------

%  Sec. 2.3.2  %%%%%%%%%%%%%%%%%%%%%%%%%%%%%%%%%% SUBSUBSECTION
\subsubsection{Radiative opacities}
%------------------------------------------------------------

In recent years, a significant progress has been made
in calculations of the radiative opacities in dense atmospheres.
We use the most recent OPAL
opacity library (Rogers et al.\ 1996). 
For iron and hydrogen, it
displays the tables of the spectral opacities, which
we convert into the Rosseland opacities $K_{\rm rad}$
using the standard frequency-averaging procedure 
(e.g., Schwarzschild 1958).
For helium, we implement the Rosseland opacities readily
available in the OPAL library.

For $(\rho,T)$ values within the table range 
($\rho < \rho_{\rm OPAL}$),
we have used bilinear interpolation of $\log K_{\rm rad}$
between the table entries in $\log\rho$ and $\log T$.
For $\rho > \rho_{\rm OPAL}$, we employ similar bilinear extrapolation
based on the nearest pairs of table entries in $\log\rho$
and $\log T$. The extrapolated values of $K_{\rm rad}$ have practically
no effect on temperature profiles, because 
at high densities the heat
transport is provided by electrons
($K_{\rm c} \ll K_{\rm rad}$).

% 2.3.2 %%%%%%%%%%%%%%%%%%%%%%%%%%%%%%%%% SUBSUBSECTION
\subsubsection{Conductive opacities}
\label{sect-conduct}
%------------------------------------------------------

Previous studies of the temperature profiles in NS envelopes
have mainly used the electron thermal conductivity
of completely ionized, strongly degenerate plasma.
In particular, one has often taken
the conductive opacities from
Urpin \& Yakovlev (1980) and Yakovlev \& Urpin (1980) (hereafter YU).
However, in a cold enough NS,
the thermal conductivity of non-degenerate or partly degenerate
electrons may become important. A comparison with the tabular data of
Hubbard \& Lampe (1969) reveals that
a straightforward extrapolation of the YU formulae
from their validity domain (fully ionized, degenerate plasma)
to the case of non-degenerate matter
may lead to an underestimation of the electron thermal conductivity
by orders of magnitude. As will be shown in Sect.~\ref{sect-res},
this may strongly affect the temperature profiles in the NS envelopes
at $T_{\rm eff}\la 10^5$~K.

In order to determine the
conductive opacities for any degeneracy,
we employ the numerical code developed recently
by Potekhin \& Yakovlev (1996) (hereafter PY) for calculating
the electron transport coefficients along magnetic fields in the
magnetized NS envelopes. The code
performs numerical thermal averaging of the effective
energy-dependent electron relaxation time
with the Fermi--Dirac distribution at any degeneracy.
This enables us to consider not only strongly degenerate,
but also mildly degenerate matter.
In addition, PY have improved the approach of YU by
a more accurate calculation of the Coulomb logarithm
for the electron-ion scattering in the liquid phase
of NS matter and by incorporating the Debye--Waller reduction
factor (e.g., Itoh et al.\ 1984)
for the electron-phonon scattering in the solid phase.

In the present article, we apply the above code for the particular
case of zero magnetic field.
For this purpose, further modifications
have been made. First,
in addition to the thermal conduction produced by the
electron-ion scattering
considered by PY, we have incorporated
also the contribution from the
electron-electron scattering in the liquid regime
(see Sect.~\ref{app-ee}). This process
can be important for light elements (H, He)
in a weakly degenerate matter.
Second, we have improved
the Coulomb logarithm which determines the
collision rate of strongly degenerate electrons with ions
(see Sect.~\ref{sect-coulog}).
Third, in the solid regime, we have used the new expression
for the electron thermal conductivity, derived recently
by Baiko \& Yakovlev (1995).
Note that at temperatures much lower than the melting
temperatures, the Coulomb scattering of electrons by 
charged impurities may become important (e.g., YU).
It depends on the impurity charge and, most important, 
on its number density $n_{\rm imp}$ 
which is generally unknown in the NS envelopes. We will present
the results obtained for pure crystals ($n_{\rm imp} \to 0$).
Our calculations including $^{56}$Ni 
impurities in iron envelopes 
show that the electron-impurity scattering may
affect the temperature profiles in the crusts of
cooling NSs, whenever the
impurity concentration is sufficiently high,
$n_{\rm imp} \ga 0.01\, n_{\rm i}$.
Finally, in the non-degenerate 
layers, we have performed the thermal averaging 
introducing modifications into the 
effective width of the Landau levels
which was used by PY to 
account for collisional 
and inelastic broadenings 
of the Landau levels in strongly degenerate matter. 
The effective width proposed by PY becomes inadequate 
in the non-degenerate regime, and we have switched it off 
by multiplying by the factor $(1+10T/T_{\rm F})^{-1}$. 

A comparison of our thermal conductivities 
with the tables of Hubbard \& Lampe (1969)
(at $\rho \la  10^5$ g cm$^{-3}$, where the tables
are adequate) shows maximum discrepancy up to
30\% for carbon and up to 70\% for lighter elements in the 
non-degenerate regime, and still smaller discrepancies
in mildly and strongly degenerate matter.
More accurate values
of the Coulomb logarithm obtained in Sect.~\ref{sect-coulog}
modify $K_{\rm c}$ by about
5--30\%, which has only a little effect on the thermal structure
of the envelope (see Sect.~\ref{sect-res}).

% Sec. 2.3.3 %%%%%%%%%%%%%%%%%%%%%%%%%%%%%%%%%% SUBSUBSECTION
\subsubsection{Coulomb logarithm}
\label{sect-coulog}
% -----------------------------------------------------------

The thermal conductivity $\kappa$ of degenerate electrons
($T < T_{\rm F}$) in a liquid or a gas of ions
with mass and charge numbers $A$ and $Z$ can be written as
\begin{equation}
      \kappa  =  { \pi^2 k_{\rm B}^2 T n_{\rm e}
                \over 3 m_\ast \nu_{\rm e}},
\label{kappa}
\end{equation}
where $m_\ast = m_{\rm e} \sqrt{1+x^2}$,
$\nu_{\rm e} = \nu_{\rm ei} + \nu_{\rm ee}$ is the effective
electron collision frequency,
$\nu_{\rm ei}$ is determined by the Coulomb electron-ion
collisions, and $\nu_{\rm ee}$ is determined by the
electron-electron collisions. New calculations and fits
of $\nu_{\rm ee}$ are presented in Sect.~\ref{app-ee}.
In this section, we examine $\nu_{\rm ei}$.
It can be expressed as (e.g., YU)
\begin{equation}
   \nu_{\rm ei}  =  {4 m_{\rm e}^\ast Z e^4 L \over 3 \pi \hbar^3},
\label{ei}
\end{equation}
where $L$ is the Coulomb logarithm to be evaluated.
It is a slowly varying function of $\rho$ and $T$.

First consider a strongly coupled Coulomb
liquid of ions, at $1 \la \Gamma \la \Gamma_{\rm m}$,
where $\Gamma_{\rm m}$ is the critical value of the ion
coupling parameter $\Gamma$ at which a Coulomb crystal melts
($\Gamma_{\rm m}$ = 172, for a classical
ion system, Nagara et al.\ 1987;
the deviations from the classical 
melting curve due to the quantum corrections 
turn out to be small at $\rho < 10^{10}$ g cm$^{-3}$ 
for carbon and heavier ions -- see Chabrier 1993).
In this case, one usually assumes that the screening of the
electron-ion Coulomb interaction is static and produced by
the ion-ion correlations. The screening is
described by the ion-ion structure
factor $S(q)$. Extensive calculations of $L$
in the Born approximation for different ion species, $Z$ and $A$,
were performed by Itoh et al.\ (1983). The authors used
the structure factors obtained in the approximation of
rigid electron background. They fitted their results by
complicated expressions. Later Yakovlev (1987) evaluated
non-Born corrections to $L$, using the weak screening
approximation.

Now we recalculate the Coulomb logarithm for strongly
degenerate electrons at $1 \la \Gamma \la \Gamma_{\rm m}$
from the expression
\begin{equation}
    L  = \int_0^{2 k_{\rm F}} \, {\rm d}q \,
        {q^3 S(q) F^2(q) R(q) \over q^4 \epsilon(q)}
  \left( 1 - {\beta^2  q^2 \over 4 k_{\rm F}^2 } \right),
\label{L}
\end{equation}
where $\beta = 
x /\sqrt{1+x^2}$,
$k_{\rm F}= p_{\rm F}/ \hbar$, $\hbar q$ is a momentum transfer
in an $ei$ collision event, 
$\epsilon(q)$ is the static longitudinal dielectric function of
electrons, $F(q)$ is the nuclear
form factor to allow for finite
nuclear size, and $R(q)$ is the non-Born correction factor 
specified below.
The dielectric function has been calculated for the electron
gas at any degeneracy. However, we have verified
that, for all the cases in the present study, this dielectric
function yields the same results
as the function obtained in the limit of strong
electron degeneracy for the relativistic electron gas (Jancovici 1962).
Thus the bulk of the calculations has been done with
this latter function.
We use the nuclear form factor $F(q)$ appropriate to
a uniform proton core of an atomic nucleus, with 
the same core radii
as in Itoh et al.\ (1983). The finite sizes of the nuclei
appear to have almost no effect on $L$, for the conditions
of study. Nevertheless we have kept $F(q)$ in our calculations.
Following Yakovlev (1987), the non-Born factor has been
taken as $R(q) = \sigma(q)/\sigma_{\rm B}(q)$,
where $\sigma(q)$ is the exact differential Coulomb scattering
cross section for a momentum transfer $q$, and
$\sigma_{\rm B}(q)$ is the cross section in the Born approximation.
The exact cross sections are taken from
Doggett \& Spencer (1956).

In our calculations from Eq.\ (\ref{L}), we have used
the ion structure factors evaluated for a {\it responsive}
electron background
including the local field corrections between electrons
(Chabrier 1990).
Our calculations extend those of Itoh et al.\ (1983)
by including the non-Born corrections and the effects of
electron response on the ion structure factors. On the other hand,
we improve the results of Yakovlev (1987) since we evaluate
the non-Born corrections
beyond the weak screening approximation.

We have evaluated $L$ from Eq.\ (\ref{L}) for chemical
elements with $1 \leq Z \leq 26$ and
for a dense grid of
$1 \leq \Gamma \leq \Gamma_{\rm m}$ and
$10^2$ g cm$^{-3} \leq \rho \leq 10^{10}$ g cm$^{-3}$
provided $T \leq T_{\rm F}$.
In the case of hydrogen, we have restricted ourselves to
$\rho \leq 10^8$ g cm$^{-3}$ since hydrogen should be completely
burnt at higher densities. For light elements,
the highest temperatures included in these data
correspond to $\Gamma =1$ and appear to be much below $T_{\rm F}$.
Accordingly, there exists a large temperature
interval in the electron degeneracy domain where
$\Gamma < 1$, the ion coupling ceases to be strong,
the ion screening cannot be treated in terms of ion-ion
correlations, and Eq.\ (\ref{L}) is invalid.
In principle, the Coulomb logarithm
for mildly-coupled ion plasma ($\Gamma \la 1$) can be
evaluated using the formalism developed by
Boerker et al.\ (1982). In order to simplify
our analysis, we will restrict ourselves to
the weak--coupling case ($\Gamma \ll 1$), in which
the ions constitute
a Boltzmann gas, the weak ion screening
is dynamical and can be described by the
dielectric function formalism,
while the electron screening remains static (e.g., Lampe 1968).
At $\Gamma \ll 1$ and $T < T_{\rm F}$, we propose the expression
\begin{eqnarray}
   L & = & { 1 \over 2} \ln \left( {1 \over b_{\rm i}} \right)
       - {1 \over 2} \, \eta \ln \left( { 1 \over \eta} \right)
+ {1 \over 2} \, (1+ \eta) \ln \left( { 1 \over 1 + \eta} \right)
\nonumber \\
      & - & {1 \over 2} \, \beta^2
       + {1 \over 2} \, \pi Z \alpha \beta,
\label{DeWitt}
\end{eqnarray}
where $ b_{\rm i} = 1/(2 k_{\rm F} r_{\rm Di}) =
1/\sqrt{3 \Gamma}$ is the
ion screening parameter in the weak coupling regime,
$r_{\rm Di}$ is the Debye radius of ions,
 $\eta = b_{\rm e}/b_{\rm i}$, $b_{\rm e}
= k_{\rm TF}/(2 k_{\rm F}) = \sqrt{ \alpha /(\pi \beta)}$
is the electron screening parameter, $k_{\rm TF}$ being 
the inverse screening length
of a charge by degenerate plasma electrons
and $\alpha$ the fine-structure constant.
The first three terms represent the Coulomb
logarithm obtained for degenerate electrons and weakly coupled
ions with the non-relativistic $ei$ scattering cross section.
The expression presented is valid for $\eta <1$; it
can be derived using the formalism
of Williams \& DeWitt (1969) to account for the
dynamic ion screening and to integrate the $ei$ collision rate
over velocities of ions.
The fourth term is the relativistic correction in the weak--coupling
limit (e.g., YU). The fifth term
is the second--Born correction in the weak--coupling limit
(Yakovlev 1987). Thus, we have supplemented our table of the
Coulomb logarithms calculated for $\Gamma \geq 1$ (see above)
with new values evaluated from Eq.\ (\ref{DeWitt})
for $\Gamma < 0.25$ and $T < T_{\rm F}$.
Analytic fits to the full set of about 31,000 values of $L$ are
presented in Sect.~\ref{app-L}. The tables of $L$ are freely
available in the electronic form.

\begin{figure}[t]
\begin{center}
\leavevmode
\epsfysize=60mm
\epsfbox[170 260 440 625]{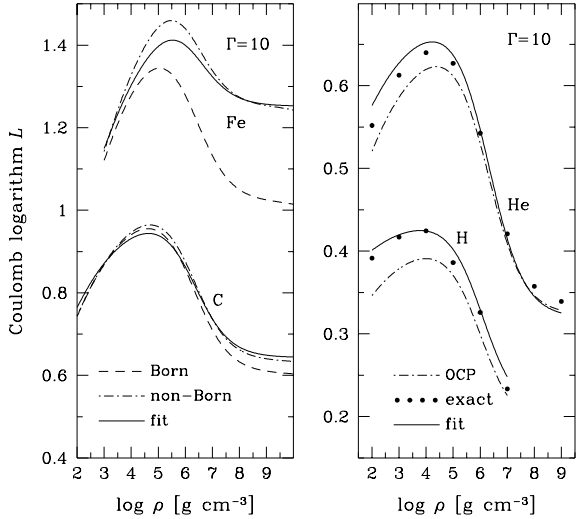}
\end{center}
\caption[]{
Coulomb logarithm $L$ vs density
for fully ionized plasmas at $\Gamma = 10$.
Left panel: Fe and C plasmas;
dash-and-dots --- present calculations,
solid line --- analytic fits (Sect.~\protect{\ref{app-L}}),
dashes --- Born approximation (Itoh et al.\ 1983).
Right panel: H and He plasmas;
dots --- present calculations,
solid line --- analytic fits,
dash-and-dots ---
ion-structure factors in the rigid electron
background (Itoh et al.\ 1983).
}
\label{fig-L}
\end{figure}

For illustration, in Fig.\ \ref{fig-L} (left panel)
we compare the Coulomb
logarithms in fully ionized Fe and C plasmas with those
calculated in the Born approximation
(Itoh et al.\ 1983). The non-Born corrections enhance
the small--angle Coulomb scattering cross section, and increase
$L$. The increase is especially pronounced for the heavier element, Fe
(exceeds 20\%), at densities $\rho \ga 10^6$ g cm$^{-3}$,
where the electrons are relativistic. The non-Born corrections
become less important with decreasing $Z$, and they are
negligible for light elements (H, He).

The inclusion of the electron response in the calculations of the
ionic structure factors affects the Coulomb logarithm
$L$ for H and He at $\Gamma \ga 1 $,
as demonstrated on the right panel of Fig.\ \ref{fig-L}.
One can observe an increase of $L$
up to 15\% for H at $\rho = 10^2$ g cm$^{-2}$ and $\Gamma = 10$
with respect to the case
of rigid electron background. Actually, the increase
grows with $\Gamma$ and reaches about 40\% at $\Gamma = 160$.
The effect is weaker for He,
and small for heavier elements (in the adopted parameter range).
Thus, for the elements heavier than He, 
we have used the structure factors
calculated in the rigid background approximation.

Note that calculated values of $L$ determine also
the electric conductivity of degenerate matter in a liquid
or gaseous ionic plasma,
$\sigma = e^2 n_{\rm e}/(m_{\rm e}^\ast \nu_{\rm ei})$.

% Fig. 2 %%%%%%%%%%%%%%%%%%%%%%%%%%%%% FIGURE

%

% Sec. 3 %%%%%%%%%%%%%%%%%%%%%%%%%%%%%%%%%%%%%%%%%%%%%%%%%%  SECTION
\section{Results and discussion}
\label{sect-res}
%---------------------------------------------------------

% Sec. 3.1. %%%%%%%%%%%%%%%%%%%%%%%%%%%%%% SUBSECTION
\subsection{Thermal structure of non-accreted envelopes}
\label{sect-non-accr}
% -----------------------------------------------------

In this section, we consider a non-accreted NS envelope
composed solely of iron, which may be at
various ionization stages.
We use the OPAL iron EOS where available,
the ideal-gas EOS with the Coulomb correction (\ref{P_C})
at high densities, and the interpolated EOS at intermediate
densities as described in Sect.~\ref{sect-interdens}.
The conduction properties are described in the mean ion approximation
with the effective ion charge $Z_{\rm eff}$ determined consistently
with the EOS (Sect.~\ref{sect-zeff}).

%% Fig. 3 %%%%%%%%%%%%%%%%%%%%%%%%%%%%%%%%%%%%% FIGURE
\begin{figure}[t]
\begin{center}
\leavevmode
\epsfysize=50mm
\epsfbox[90 260 475 580]{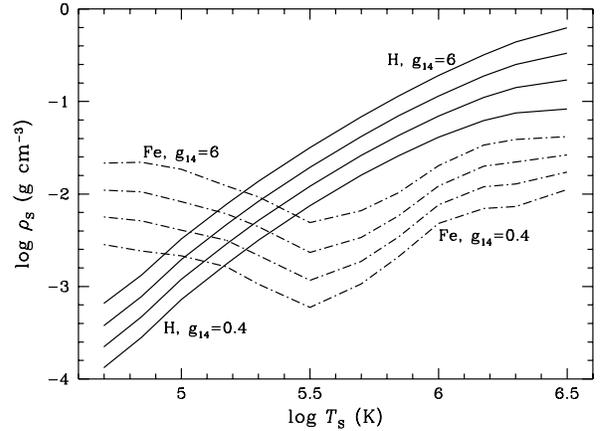}
\end{center}
\caption[]{
Boundary conditions for the thermal structure
equation (\protect\ref{th-str}).
Surface density is obtained from Eq.~(\protect\ref{Ps})
as a function of surface temperature,
for gravities $g_{14}$=0.4, 1, 2.43, 6.
Dash-dotted lines correspond to a non-accreted matter
(Fe opacities), and solid lines to an accreted matter
(H opacities).
}
\label{fig-rho-s}
\end{figure}
%
%% Fig. 4 %%%%%%%%%%%%%%%%%%%%%%%%%%%%%%%%%%%%%%% FIGURE
\begin{figure}[t]
\begin{center}
\leavevmode
\epsfysize=100mm
\epsfbox[165 185 430 650]{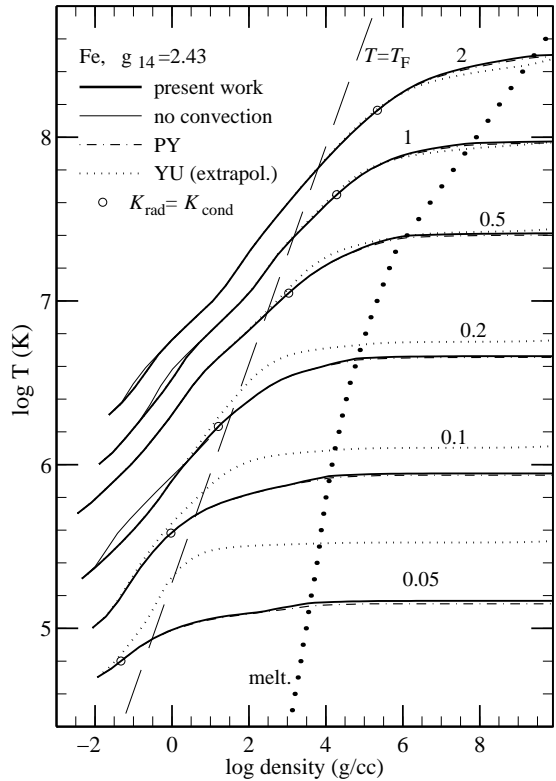}
\end{center}
\caption[]{
Temperature profiles in a non-accreted NS
(heavy solid lines) compared with various approximations:
after Yakovlev \& Urpin (1980) (YU) by extrapolating
the conductive opacities to $T > T_{\rm F}$;
after Potekhin \& Yakovlev (1996) (PY); and with
neglected convection (thin solid lines).
The curves are labeled by the values 
of $T_{\rm eff}/10^6 \, {\rm K}$.
Circles show the points, where the radiative
opacity equals the conductive
one; heavy dots show the melting curve
$\Gamma=172$; long dashes display 
the degeneracy curve, $T = T_{\rm F}$.
}
\label{fig-prfe}
\end{figure}
%
%% Fig. 5 %%%%%%%%%%%%%%%%%%%%%%%%%%%%%%%%%%%% FIGURE
\begin{figure}[t]
\begin{center}
\leavevmode
\epsfysize=100mm
\epsfbox[130 170 455 660]{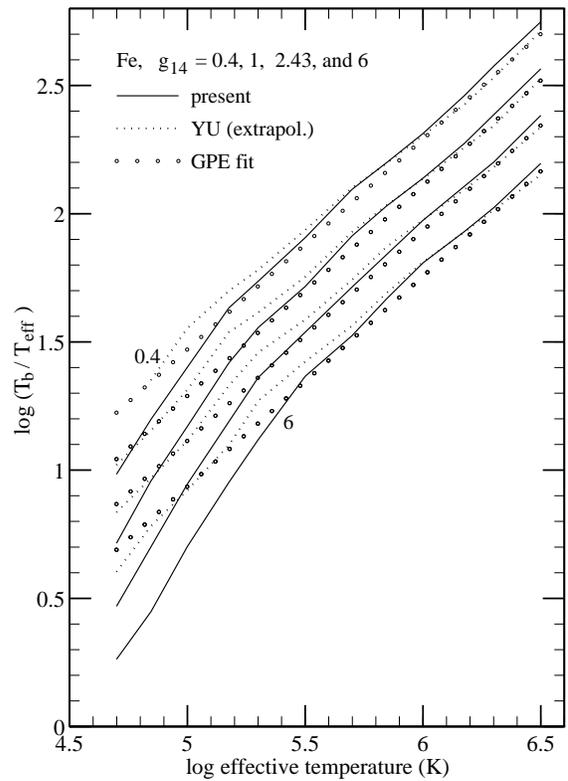}
\end{center}
\caption[]{
Temperature increase through
non-accreted NS envelopes with different
values of the surface gravity 
as compared to Eq.~(\protect\ref{GPEfit}).
}
\label{fig-link-Fe}
\end{figure}

Figure \ref{fig-prfe} presents density
dependence of the temperature in the NS 
envelope at various $T_{\rm eff}$.
The integration of Eq.~(\ref{th-str}) has been started with
the surface density $\rho_{\rm s}$,
shown in Fig.~\ref{fig-rho-s},
and terminated at $\rho_{\rm b}=10^{10}$ g cm$^{-3}$.
The value $g_{14}=2.43$ chosen in Fig.~\ref{fig-prfe}
corresponds to a
NS with a radius $R=10$ km and a mass $M=1.4\,M_\odot$.
Circles on the curves are the points of equal
radiative and conductive opacities.
The thermal flux is mainly carried by
radiation at lower densities
and by electrons at higher densities.
We also show the electron degeneracy curve and the melting curve.

In a wide range of $T_{\rm eff}$, some layers
(Fig.~\ref{fig-prfe}) turn out to be convectively unstable
(Zavlin et al.\ 1996, Rajagopal \& Romani 1996).
In these layers, the energy transport is dominated
by convection rather than by electron or radiative conduction.
We describe the convective energy flux
in the adiabatic approximation (Schwarzschild 1958),
which assumes that the convective heat transport is much
more efficient than other mechanisms.
Accordingly, in the convective zones, we replace
the temperature gradient given by Eq.~(\ref{th-str})
by the adiabatic gradient provided by the EOS tables.

In order to check the effect of convection, we have performed
also calculations with ``frozen'' convection
(i.e., neglecting the convection effect).
This is the extreme case opposite to the adiabatic one. 
In this approximation, 
we obtain (Fig.~\ref{fig-prfe}) slightly higher temperatures
inside the convective part of the atmosphere.
The atmospheric temperature profiles derived
by Zavlin et al.\ (1996), who solved numerically the 
radiative transfer equation at moderate optical depths 
and described the convection using
the mixing-length theory, lie between our two extremes.
In deeper layers, the two extreme profiles tend to merge,
as clearly seen in Fig.~\ref{fig-prfe}.
This merging occurs because the 
factor $K/T^{4}$ on the r.h.s.\ of Eq.~(\ref{th-str}) 
rapidly decreases, with increasing $T$, 
and thus ensures the isothermal profile at higher $T$ 
behind the convective zone (at fixed $T_{\rm eff}$). 
Thus, the thermal structure of
the NS envelope at $\rho\ga 10$ g cm$^{-3}$
is practically unaffected by the convection.

The dot-dashed curves in Fig.~\ref{fig-prfe} are obtained
with $K_{\rm c}$ calculated
according to PY (Sect.\ \ref{sect-conduct}). These curves
almost coincide with the solid lines,
calculated more accurately. This confirms
the validity of the PY approximations for the non-accreted envelopes.

If $T_{\rm eff} \la 2 \times 10^5$~K,
the electron conduction becomes important
not only in degenerate matter, but also
in a non-degenerate region (Fig.~\ref{fig-prfe}).
Then the straight-line logarithmic extrapolation
of the YU conductive opacities into the non-degenerate regions
leads to a significant overestimation
of the internal temperature (dotted curves).

This effect is seen also in
Fig.~\ref{fig-link-Fe}, that shows
$\log(T_{\rm b}/T_{\rm eff})$ as a function of $\log T_{\rm eff}$
for four values of the surface gravity.
If $\log T_{\rm s}\,[\mbox{K}]>5.5$,
the well-known simple fit of GPE,
\begin{equation}
   T_{\rm b6}=128.8\,(T_{\rm s6}^4/g_{14})^{0.455},
\label{GPEfit}
\end{equation}
is rather accurate, but at lower $T_{\rm s}$ the
deviations from this fit are
quite pronounced. Nevertheless, the scaling law
$
   T_{\rm b}=f(T_{\rm s}^4/g)
$
holds with a good accuracy in the whole temperature--gravity range
presented in this figure, 
except for the lowest $T_{\rm b}$ values.
The fit which reproduces these new features is presented
in Sect.~\ref{app-link}.

% Sec. 3.2. %%%%%%%%%%%%%%%%%%%%%%%%%%%%%%%%%%%%%%%%%%% SUBSECTION
\subsection{Thermal structure of accreted envelopes}
\label{sect-accr}
%-----------------------------------------------------------------

%%Fig. 6 %%%%%%%%%%%%%%%%%%%%%%%%%%%%%%%%%%%%% FIGURE
\begin{figure}[t]
\begin{center}
\leavevmode
\epsfysize=105mm
\epsfbox[110 145 480 675]{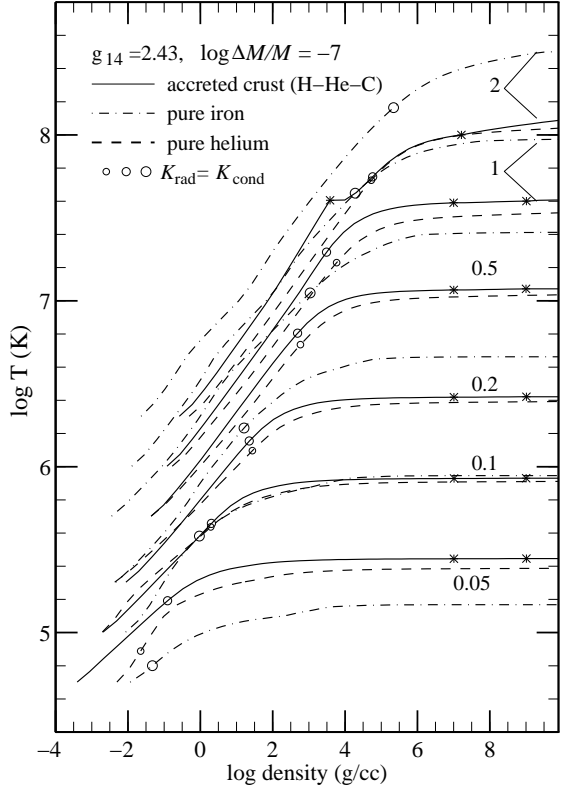}
\end{center}
\caption[]{
Temperature profiles in an accreted NS
(solid lines) as compared to
those in NS envelopes
composed of pure iron
(dot-dashed lines) and of pure helium (dashed line).
The curves are labeled by
$T_{\rm eff}/10^6 \, {\rm K}$;
full circles separate the regions of
radiation and electron conduction;
asterisks indicate the H/He (left) and He/C (right) interfaces.
}
\label{fig-prc}
\end{figure}

The chemical composition of an accreted envelope depends
on the composition of the infalling matter and
on the nuclear transformations (thermo- or pycnonuclear
burning, beta-captures) which occur while the matter sinks
within the NS. The structure and stability of
the accreted envelopes have been considered in number of
works devoted to bursting NSs (see, e.g., Ayasli \& Joss 1982,
Paczy\'{n}ski 1983, Ergma 1986, 
Miralda-Escud\'{e} et al.\ 1990,
Blaes et al.\ 1992, and references therein).
However, most of the work focused either
on high accretion rates or on slowly accreting old NSs
whose internal temperature is solely determined by the accretion.
On the contrary, we are mainly interested in
not too old, cooling NSs with thin accreted envelopes,
in which the heat release due to
the nuclear transformations has no effect on the thermal
structure.

Let us assume that an accreted envelope consists of
shells with different chemical compositions.
For an accretion of H/He, it is reasonable to expect that
the outermost accreted layers are built of
pure H, owing to the strong gravitational
stratification (Alcock \& Illarionov 1980),
whereas He may exist in the deeper layers.
The accretion is accompanied by
the nuclear transformations of H/He into heavier
elements. The details of these transformations are still uncertain,
and we consider several models of accreted envelopes.

As a first example, we take the typical
temperatures and densities of the
nuclear burning from Ergma (1986) and assume the burning to
be non-explosive.
If $T_{\rm eff}$ is high and the accreted hydrogen layer reaches
the region with $T\sim 4 \times 10^7$~K, then 
H burns efficiently into He in the
thermonuclear regime.
If temperature is lower and the hydrogen layer reaches the density
$\rho \sim 10^7$ g cm$^{-3}$, a pycnonuclear 
burning of H into He proceeds. 
At higher $\rho \sim 10^9$ g cm$^{-3}$
and/or $T\sim 10^8$~K, helium burns into carbon 
(Schramm et al.\ 1992).
The pycnonuclear carbon burning occurs
at $\rho\sim10^{10}$ g cm$^{-3}$ (e.g., Yakovlev 1994),
which is already inside the isothermal region of the envelope.

Figure \ref{fig-prc} displays the thermal structure
of a fully accreted envelope
(the accreted mass $\Delta M \sim 10^{-7} M$  fills
the layers up to $\rho \simeq \rho_{\rm b}$)
of a NS with the surface gravity $g_{\rm s} = 2.43 \times 10^{14}$
cm s$^{-2}$. The outer, intermediate, and inner shells
of this envelope, separated by asterisks,
are composed of H, He, and C, respectively.

One can observe significant differences from the thermal structure
of a non-accreted Fe envelope, as explained below. For a not too cold NS
($T_{\rm eff} \ga 10^5$~K), the main temperature gradient
occurs in
a layer of degenerate electron gas with ions
in the liquid state, where the thermal conduction is
mostly provided by the electrons due to the electron-ion
collisions. The heavier the element, the smaller the thermal
conductivity, and the steeper is the temperature growth
inside the star. However, with decreasing $T_{\rm eff}$,
the width of the heat--insulating degenerate layer
becomes smaller, and the effect is less pronounced.
In a cooler NS ($T_{\rm eff} \la 10^5$~K), the main temperature
gradient shifts into the NS atmosphere (to the optical depths 
$\sim 1$). For heavier elements,
the atmospheric layers appear generally denser
(at the same $T_{\rm eff}$, see Fig.~\ref{fig-rho-s}).
Then the internal temperature gradient is weaker
and the temperature grows slower inside the star.
Thus, a not too cold accreted envelope is more heat--transparent
than the iron envelope, whereas a cold accreted envelope is
less heat-transparent. As seen from
Fig.\ \ref{fig-rho-s}, the effective surface temperature
which separates these two regimes is almost independent of
the surface gravity.

%% Fig. 7 %%%%%%%%%%%%%%%%%%%%%%%%%%%%%%%%%%%%%%%% FIGURE
\begin{figure}[t]
\begin{center}
\leavevmode
\epsfysize=60mm
\epsfbox[180 260 440 625]{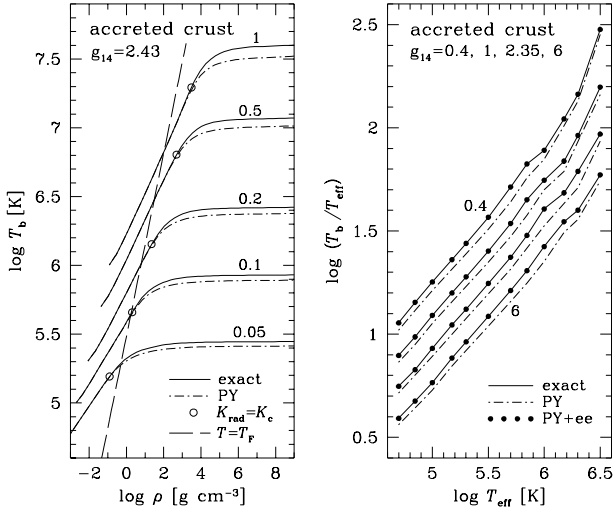}
\end{center}
\caption[]{
Temperature increase through
accreted NS envelopes for selected values of the 
surface gravity calculated using the full code
(solid lines), PY-code (dot-dash lines), and
PY-code supplemented by thermal conduction
due to electron-electron collisions (dots).
}
\label{fig-PY}
\end{figure}

Figure \ref{fig-PY} demonstrates the effect of the  
improvements in the conductive opacities.
The present results are compared with those in which
the electron thermal conductivity is calculated with the PY code.
Although the difference is not too large, it is higher
than for the non-accreted Fe envelopes (Fig.~\ref{fig-prfe}).
The difference comes mostly from the electron-electron 
collisions (neglected in PY). Dots in Fig.\ \ref{fig-PY} show
the results obtained with the PY code supplemented by
the contribution from the electron-electron collisions
to the thermal conduction (Sect.~\ref{app-ee}).
The difference from the accurate results practically disappears.

Now let us analyze the sensitivity of the
temperature profiles to the positions of the
H/He and He/C interfaces.
These interfaces, indicated by asterisks in Fig.~\ref{fig-prc},
are not well theoretically established.
For example, the densities of nuclear H and He burning
presented by Iben (1974) (see also Kippenhahn \& Weigert 1990)
are much lower than those given by Ergma (1986).
We have significantly
shifted the interfaces according to the results of Iben (1974),
but the $T_{\rm eff}$--$T_{\rm b}$ relationship
remains practically unchanged.
We have also calculated the temperature
profiles for a purely helium NS envelope.
These profiles are slightly different from those
in a burning accreted matter described above. However, the
$T_{\rm eff}$--$T_{\rm b}$ relationships remain 
practically the same.

%%Fig. 8 %%%%%%%%%%%%%%%%%%%%%%%%%%%%%%%%%%%% FIGURE
\begin{figure}[t]
\begin{center}
\leavevmode
\epsfysize=100mm
\epsfbox[130 170 455 660]{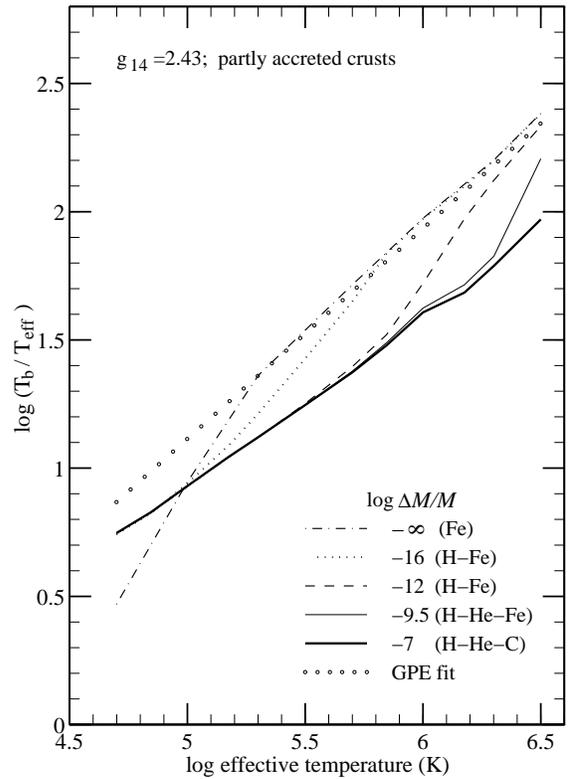}
\end{center}
\caption[]{
Temperature increase through partly accreted NS crusts.}
\label{fig-link-a}
\end{figure}

Recently the thermal structure of old,
slowly accreting NSs has been examined in several articles
(see Miralda-Escud\'{e} et al.\ 1990, Blaes et al.\ 1992,
and references therein). In these studies,
the internal stellar temperature is assumed to be
determined by the accretion rate.
The composition of the accreted envelopes appears to be
similar to the ones considered above except for the
innermost layers. For instance, Miralda-Escud\'{e} et al.\ (1990)
assume that He transforms directly into Fe while
Blaes et al.\ (1992) state that the burning of
He into C at $\rho\sim 10^9$ g cm$^{-3}$ is followed
by the carbon transformation through the rapid
$^{12}$C$(\alpha,\gamma)^{16}$O process. Accordingly,
we have recalculated the temperature profiles
replacing $^{12}$C by $^{56}$Fe or
$^{16}$O and found only very small differences.
The insensitivity of the temperature profiles
to the composition of the innermost layers of the
envelope is quite natural, because
these layers are nearly isothermal.

Summarizing the above discussion
we can conclude that the $T_{\rm eff}$--$T_{\rm b}$
relationship is remarkably insensitive
to the parameters of nuclear burning and to the chemical
composition of accreted matter (H and/or He) but differs
from the one in a non-accreted NS. This conclusion will not change
if H burning is explosive (see, e.g., Paczy\'{n}ski 1983,
Miralda-Escud\'{e} et al.\ 1990).
During an outburst, the H-shell burns
into He, without any noticeable effect on the
$T_{\rm eff}$--$T_{\rm b}$ relation. According to
Miralda-Escud\'{e} et al.\ (1990) helium burning
in slowly accreting NSs
is unstable; it produces nuclear outbursts
with burning of all accreted material.
Nevertheless, the helium layer should reach a very high density,
$\rho \sim 10^9$ g cm$^{-3}$ to trigger the instability
in a not too hot NS.
If the same is true for the conditions of our interest,
our results are valid during a
long period of time until the instability is reached.
However, according to Paczy\'{n}ski (1983),
the presence of the heat outflow from a not too cold,
cooling NS essentially stabilizes
nuclear burning in accreted envelope, and the He burning is
mostly non-explosive for the parameters we are interested in.

Figure~\ref{fig-link-a} shows
$\log(T_{\rm b}/T_{\rm eff})$ as a function of $\log T_{\rm eff}$
for various amounts of accreted matter.
The dot-dashed line represents $\log(T_{\rm b}/T_{\rm eff})$ for
a non-accreted envelope from Fig.~\ref{fig-link-Fe}.
This function is strongly affected by the accretion.
Even a thin hydrogen (or He) mantle of mass $\Delta M=10^{-16}M$,
which extends only to $\rho \sim 10^3$ g cm$^{-3}$,
modifies strongly the $T_{\rm b}$--$T_{\rm eff}$ relation.
Analytical fits of $T_{\rm eff}$
as a function of the internal temperature $T_{\rm b}$,
surface gravity $g$, and accreted mass $\Delta M$
are given in Sect.~\ref{app-link}.

% Sec. 3.3  %%%%%%%%%%%%%%%%%%%%%%%%%%%%%%%%%%%%%%%%%%%% SUBSECTION
\subsection{Effect of accreted envelope on cooling}
\label{sect-cooling}
%------------------------------------------------------------------

Consider briefly the effects of the possible
presence of accreted matter in the NS envelopes on the thermal history
of NSs. For this purpose, we simulate the cooling of NSs using
two NS models. We use the cooling code of Gnedin \& Yakovlev
(1993) based on the approximation of isothermal NS interior.
The code follows accurately the cooling of a not too young
NS whose thermal relaxation
is already over ($t \ga 10^2$ yrs, 
see Nomoto \& Tsuruta 1987, Umeda et al.\ 1994).
We adopt a moderately stiff EOS of Prakash et al.\ (1988)
to describe matter in the NS cores.
This matter is assumed to consist of neutrons, protons, and electrons
(no exotic cooling agents such as quarks or kaon condensates).
The maximum NS mass, for this EOS, is about 1.7 $M_\odot$.
We examine two NS models: (1) $M=1.30 \, M_\odot$,
$R = 11.72$ km, the central density
$\rho_{\rm c} = 1.12 \times 10^{15}$ g cm$^{-3}$;
(2) $M=1.44 \, M_\odot$, $R = 11.35$ km,
$\rho_{\rm c} = 1.37 \times 10^{15}$ g cm$^{-3}$.

The first model is a typical example of the {\it standard} cooling.
The central density is insufficient to
switch on the most powerful neutrino cooling
due to the direct Urca process (Lattimer et al.\ 1991).
In this case the main neutrino energy loss mechanisms in the NS cores
are the modified Urca reactions (neutron and proton branches)
and the nucleon-nucleon bremsstrahlung
(Friman \& Maxwell 1979, Yakovlev \& Levenfish 1995).

Our second NS model is an example of the {\it rapid} neutrino cooling.
In this model, the central density slightly exceeds
the threshold density $\rho_{\rm cr} = 1.30 \times
10^{15}$ g cm$^{-3}$ of the direct Urca process (for the given EOS).
Then the NS has a small central kernel 
(of radius 2.32 km and mass 0.035 $M_\odot$) where the direct Urca
process operates, and the neutrino luminosity exceeds
the standard one by several orders of magnitude.

\begin{figure}[t]
\begin{center}
\leavevmode
\epsfysize=90mm
\epsfbox[100 170 450 650]{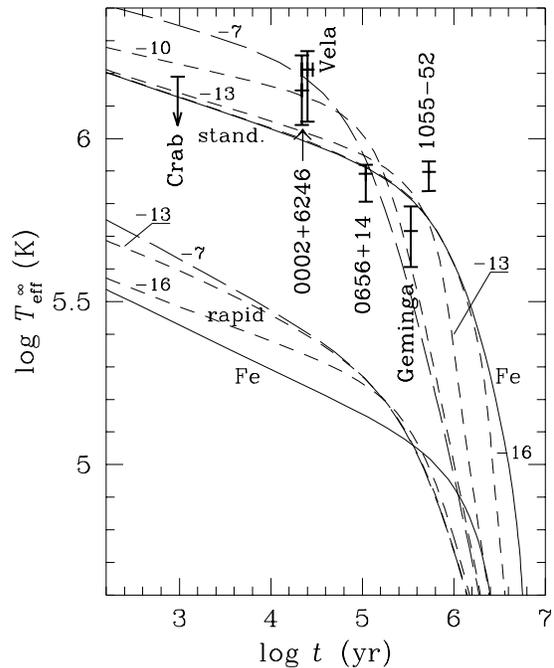}
\end{center}
\caption[]{
Standard and rapid cooling of NSs with
non-accreted (Fe) and partly and fully
accreted envelopes, compared with observations of thermal
radiation from 6 pulsars (see text).
The accreted mass is assumed to be accumulated during the first 100 yrs;
numbers along the curves denote the values of $\log\Delta M/M$.
}
\label{fig-cooling}
\end{figure}

All the neutrino energy loss rates included in the calculations
are described by Levenfish \& Yakovlev (1996).
To simplify our analysis, we assume that the NS cores are
non-superfluid, and there is no internal reheating
(e.g., energy dissipation due to differential rotation).
The results are presented in Fig.\ \ref{fig-cooling}
in the form of the cooling curves --
effective temperatures $T_{\rm eff}^{\infty}$ as measured
by a distant observer (see Sect.\ \ref{sect-th-str-eq})
versus NS age $t$.

Figure \ref{fig-cooling} presents the cooling curves
under the assumption that all the accreted material
is accumulated at the NS surface during
the NS birth. The accretion can occur, for instance,
at the post-supernova
stage (e.g., Chevalier 1989, 1996, Brown \& Weingartner 1994)
especially if the pulsar activity is suppressed initially
by burying the pulsar magnetic field under the fallen-back matter
(Muslimov \& Page 1995).
We show the standard and rapid neutrino
cooling for different amount of accreted mass $\Delta M$.
The fraction of accreted mass $\Delta M/M$
varies from 0 (non-accreted NSs) to $\sim 10^{-7}$
(fully accreted NS envelope). It is evident that
further increase of $\Delta M$ does not affect the cooling.

The change of slopes of the cooling curves at $t= 10^5$--$10^6$ yrs
reflects the change of the cooling regime. Initially,
a NS is at the neutrino cooling stage. It
cools mainly via the neutrino emission; the internal
stellar temperature is ruled by this emission
and is thus independent of the thermal insulation
of the NS envelope. The surface photon emission
is determined by the $T_{\rm b}$--$T_{\rm eff}$ relation.
Since the accreted envelopes of not too cold NSs are more
thermally transparent (Sect.\ \ref{sect-accr}), the surface
temperature of an accreted NS is noticeably higher than that
of a non-accreted NS. One can see that
even a very small fraction of accreted matter, such as
$\Delta M/M \sim 10^{-16}$, may change appreciably
the thermal history of the star.
The colder the star, the smaller is the fraction
of accreted material which yields the same cooling curve
as the fully accreted NS envelope.
The effect is naturally more pronounced for the rapid
cooling (for colder NSs).

If $t \ga 10^5$--$10^6$ yrs, the neutrino luminosity
decreases, and the NS cools mainly via the surface emission of
photons (the photon cooling stage).
Since the envelopes of the accreting (and not too cold) 
NSs have lower thermal insulation, these stars cool now 
faster than the non-accreted ones (Fig.\ \ref{fig-cooling}). 

%%%%%%%%%%%%%%%%%%%%%%%%%%%%%%%%%%%%%%%%%%% FIGURE 9

The change of the thermal
history of a NS by the accreted material should be taken into account
in theoretical interpretation of observational data
on thermal radiation from NSs. For illustration, 
in Fig.~\ref{fig-cooling} 
we present the effective temperatures with $1\sigma$ 
confidence intervals 
of the pulsars 
0833--45 Vela (\"{O}gelman et al.\ 1993, \"{O}gelman 1995),
0630+18 Geminga (Halpern \& Ruderman 1993),
0656+14 and 1055--52 (Greiveldinger et al.\ 1996), 
RXJ 0002+6246 (Hailey \& Craig 1995), 
and the $3\sigma$ upper limit for the Crab pulsar (PSR 0531+21,
Becker \& Aschenbach 1995). These data have been deduced from
the fits of spectral observations of the {\it ROSAT}
X-ray observatory. For Geminga, PSR 1055--52, and
PSR 0656+14, we plot the values of $T_{\rm eff}$
obtained for ``soft'' components
of the two- and three-component X-ray spectra
models, since it is the soft component
that is associated with the thermal flux from the
stellar interior. For Vela, we extend downward the confidence 
interval obtained by \"{O}gelman et al.\ (1993) according 
to the more recent spectral fitting of \"{O}gelman (1995), 
that gives a more plausible NS radius than the previous fit. 

The age of the Crab pulsar in Fig.~\ref{fig-cooling} is its true 
age known from the chronicles, the age of Vela is its
possible true age determined
recently by Lyne et al.\ (1996), 
the age of RXJ 0002+6246 is an estimate for the 
related supernova remnant G\,117.7+06, 
while for the three other pulsars we adopt
their characteristic ages.

We plot the values of $T_{\rm eff}$
derived by the authors by fitting the 
assumed blackbody thermal spectrum to the measured 
X-ray spectrum. 
The alternative often used is to derive $T_{\rm eff}$
from the apparent luminosity in the whole spectral range observed. 
However, this latter way is less certain, because the inferred 
temperature strongly depends on  
the assumed NS distance (usually poorly known) 
and model-dependent radius. 

Note that the blackbody fits do not take into account 
the influence of the magnetized atmosphere.
The fits for PSR 0656+14 (Anderson et al.\ 1993), 
Geminga (Meyer et al.\ 1994), and Vela (Page et al.\ 1996), 
which employ the models
of magnetized atmospheres, have resulted in a few times lower
$T_{\rm eff}$ than the blackbody models.
Our justification of using here the blackbody data is that
the so far developed models of magnetized atmospheres 
do not take proper account of absorption by 
neutral atoms. 
The neutral fraction can be significant at $T \sim 10^5-10^6$ K,
because the magnetic fields $\sim 10^{12}$~G increase 
the ground-state binding energy by about one order
of magnitude. The opacities of hydrogen in such strong fields 
are still under investigation (see 
Pavlov \& Potekhin (1995) and references therein).

The magnetic fields affect also the thermal insulation
of the NS envelopes and, therefore, the cooling
(e.g., Yakovlev \& Kaminker 1994). We do not take this
effect into account in the present article.
As suggested by Hernquist (1985) and 
shown recently by Shibanov \& Yakovlev (1996),
the dipole surface magnetic fields $ \la 3 \times 10^{12}$~G
affect the cooling much more weakly than predicted by 
the studies of simplified magnetic field geometries 
(e.g., Van Riper 1988, 1991).

As seen from Fig.\ \ref{fig-cooling},  
for the pulsars Vela, Geminga, and RXJ 0002+6246, 
the presence of accreted matter would simplify
the explanation of the cooling
by the standard neutrino emission without
invoking superfluidity of NS matter, as favoured in previous work
(Page 1994, Levenfish \& Yakovlev 1996). 
Interestingly, the accretion rate of Vela 
estimated by Morley (1996), multiplied by its age, 
yields $\log\Delta M/M_\odot\approx -9.5$, that just fits the 
middle of the error bar in Fig.~\ref{fig-cooling}. 
For PSR 0656+14, the scenarios with and without the accretion 
are equally consistent with the result of the fitting 
shown if Fig.~\ref{fig-cooling}, especially if we assume that
its braking index is somewhat lower than 3
(as happens for all 4 pulsars with
known braking indices, see Lyne et al.\ 1996)
and therefore its true age is somewhat higher than 
its characteristic age. The relatively high temperature of 
PSR 1055--52 cannot be explained by the considered models 
of the cooling, that probably suggests some reheating process 
for this relatively old pulsar. 

We have used, for illustration,
the standard NS models and simplified
blackbody fits to the observed spectra. More sophisticated
models and fits can also gain from the assumption 
of accreted envelopes. For instance, Page (1996) considers 
the cooling of NSs with kaon condensed cores 
and reheating in the outer crusts, 
using our models of the envelopes. 

% Sec. 4   %%%%%%%%%%%%%%%%%%%%%%%%%%%%%%%%%%%%%%%%%%%%%%% SECTION
\section{Conclusions}
% --------------------------------------------------------------

We have considered thermal structure and evolution of
NSs whose envelopes are composed of non-accreted or
accreted matter. We have used new, state-of-the-art calculations of EOS and
opacities of NS envelopes, described in Sect. 2.
In particular, we have recalculated
(Sects.\ \ref{sect-coulog}, \ref{app-ee}, and \ref{app-L}) 
the electron-electron and
electron-ion collision frequencies,
that determine the electron thermal conductivity,
for a wide range of densities and temperatures of a degenerate electron gas
and ionic liquid plasmas.

Using this new physics input, we have calculated 
(Sects.\ \ref{sect-non-accr}, \ref{sect-accr}) 
the temperature profiles in the envelopes
of non-accreted and partly accreted NSs and
obtained the relationships between the
internal and effective temperatures of NSs, $T_{\rm b}$ and
$T_{\rm eff}$. These relationships are 
important for simulating the NS cooling; they are fitted
by simple analytic expressions in Sect.~\ref{app-link}.
They appear to be very sensitive to the presence of accreted matter
in the NS envelope; even a small amount of the accreted matter,
$\Delta M \sim 10^{-16} \, M_\odot$, can
reduce substantially the thermal insulation
of the envelopes.
For a non-accreting NS, our relationship
between $T_{\rm b}$ and $T_{\rm eff}$ extends the well--known
result of GPE to lower temperatures
(down to $T_{\rm eff} \sim 50\,000$ K).

In Sect.\ \ref{sect-cooling}, 
we have examined briefly the effect of the possible
presence of accreted matter on the NS cooling. We show
that the accreted matter may increase the surface temperature
(photon thermal luminosity)
at the neutrino cooling stage, and decrease them 
at the subsequent photon cooling stage,
as compared to the NSs without accreted envelopes. 
We have shown that these results
can be important for a proper interpretation of
observed thermal radiation from NSs. 
In particular, the presence of accreted matter
facilitates the explanation of recent observational results 
concerning the pulsars Vela and Geminga, and RXJ 0002+6246, 
in the framework of the standard neutrino emission model
(without exotic matter, superfluidity, or direct Urca processes). 

In this paper, we have neglected effects of magnetic fields 
on the EOS and the thermal conduction of matter, which 
can be significant (e.g., Yakovlev \& Kaminker 1994).
They do deserve further studies
using improved EOS and thermal conductivities
of magnetized NS envelope (e.g., PY) and
improved radiative opacities of magnetized NS atmosphere
(e.g., Pavlov \& Potekhin 1995), with allowance for
the possible presence of light elements in the surface
layers.

Finally, it is worth noting that the physics input used
in the present calculations can be applied to a variety of
other astrophysical problems concerning dense stellar matter, e.g.
the thermal structure
and bursting activity of X-ray bursters 
(see, e.g.,
Miralda-Escud\'{e} et al.\ 1990 and references therein)
and the cooling of white dwarfs (Segretain et al.\ 1994).

\begin{acknowledgements}
We are grateful to H.\ DeWitt for pointing out
the article of Boerker et al. (1982), to
Dany Page and Yu.A.\ Shibanov for fruitful 
discussions, and
to D.\ Page for reading the manuscript and making the
most useful critical remarks. 
This work was supported in part by the
RBRF grant No.\ 96-02-16870a, INTAS grant No.\ 94-3834, 
and DFG--RBRF grant No. 96-02-00177G.
AYP gratefully acknowledges the hospitality 
of the theoretical astrophysics group of the 
Ecole Normale Sup\'erieure de Lyon. 
\end{acknowledgements}
%%%%%%%%%%%%%%%%%%%%%%%%%%%%%%%%%%%%%%%%%%%%%%%%%%%%%%%%%  APPENDIX
\appendix
\addtocounter{section}{1}
\section*{Appendix: fitting formulae}
\subsection{Electron-electron collisions}
\label{app-ee}

The effective
collision frequency $\nu_{\rm ee}$ of non-relativistic
degenerate electrons ($x \ll 1$, $T < T_{\rm F}$) was analyzed
by Lampe (1968) using the formalism
of the dynamic screening of the electron--electron interaction.
The expression of $\nu_{\rm ee}$ for the relativistic
degenerate electrons at $T \ll T_{\rm pe}$ was
obtained by Flowers \& Itoh (1976). Here,
$T_{\rm pe}$ is the electron plasma temperature
determined by the electron plasma frequency
$\omega_{\rm pe}$,
\begin{equation}
     T_{\rm pe}= {\hbar \omega_{\rm pe} / k_{\rm B}},\;\;\;
     \omega_{\rm pe} =
     \sqrt{ { 4 \pi e^2 n_{\rm e} / m_{\rm e}^\ast}},
\end{equation}
$m_{\rm e}^\ast=m_{\rm e} \sqrt{1+x^2}$.
The degeneracy temperature $T_{\rm F}$ in the NS envelopes is
typically higher than $T_{\rm pe}$. 
Urpin \& Yakovlev (1980) extended
the results of Flowers \& Itoh (1976) to higher
temperatures, $T < T_{\rm F}$. In the approximation of
static electron screening of the Coulomb interaction,
Urpin \& Yakovlev (1980) obtained
\begin{equation}
     \nu_{\rm ee}  =  {3 \alpha^2 \, (k_{\rm B} T)^2
          \over 2 \pi^3 \,
          \hbar m_{\rm e}^\ast c^2 \, b_{\rm e}^{3/2}} \, J(y),
\label{ee}
\end{equation}
where $y= \sqrt{3} \, T_{\rm pe}/T$, $\alpha=e^2/(\hbar c)$,
$\beta = x / \sqrt{1+x^2}$ and $b_{\rm e}= \alpha/(\pi \beta)$
(see Eq.\ (\ref{DeWitt})).
Now it is sufficient to calculate
the function $J(y)$, presented by Urpin and
Yakovlev (1980) as a 2D integral which depends parametrically on
the relativistic parameter $x$.
Lampe (1968) analyzed this function
in the static screening approximation at $x \ll 1$.
The asymptotes of $J$ were obtained by Lampe (1968) for $x \ll 1$ at
$y \ll 1$ and $y \gg 1$, by Flowers \& Itoh (1976)
for $y \gg 1$ at any $x_{\rm}$,
and  by Urpin \& Yakovlev (1980) for $y \ll 1$ and $x \gg 1$. 
Timmes (1992) calculated $J(y)$
numerically in the limit of $x \gg 1$. However,
the unified expression of $J(y)$
at $T<T_{\rm F}$ valid equally for relativistic and
non-relativistic electrons has been absent.
Note that the fit expression of $J(y)$
obtained by Timmes (1992) (his Eq.~(10)) 
is valid only at $y < 10^3$.

We have calculated
$J$ numerically for a dense grid of $x$ and $y$ in the
intervals $0.01 \leq x \leq 100$ and
$0.1 \leq y \leq 100$. The results are
fitted by the expression:
\begin{eqnarray}
 J & = & \left( 1 + {6 \over 5 x^2} +
     {2 \over 5 x^4} \right) \,
     \left[ {y^3 \over 3 (1+ 0.07414 \, y)^3} \right.
\nonumber \\
   & \times &
    \left.   \ln \left( {2.810 -0.810 \beta^2 + y \over y} \right)
  + {\pi^5 \over 6} \, {y^4 \over (13.91 +y)^4} \right]
\label{Jfit}
\end{eqnarray}
which reproduces also all the asymptotic limits
mentioned above. The mean
error of the fits is 3.7\%, and the maximum error of
11\% takes place at $x=1$ and $y=0.1$.

%%%%%%%%%%%%%%%%%%%%%%%%%%%%%%%%%%%%%%%%%%%%%   APPENDIX B
\subsection{Coulomb logarithms}
\label{app-L}
The tables of the Coulomb logarithms calculated from
Eqs.\ (\ref{L}) and (\ref{DeWitt}) as described in
Sect.\ \ref{sect-coulog} are fitted by
\begin{eqnarray}
     L  & = & { 1 \over 2} \, (1 + c_5 \, b)
               \ln \left( { 1 + c_7 b_{\rm e} \over b } \right) - c_4
\nonumber   \\
        & - & {1 \over 2} \, \beta^2 ( 1 - c_6 b) +
               {1 \over 2} \pi Z { \alpha \beta},
\label{Fit}
\end{eqnarray}
where $b= b_{\rm e} + b_{\rm i}$ and
\begin{eqnarray}
    b_{\rm i}  =  \left( { 3 \over 2 \pi Z } \right)^{2/3}
      \left( 1.5 c_1 +
        { c_3 \sqrt{Z} + c_2 \beta^2 \over \sqrt{\Gamma Z}}
     +  { 3 \over \Gamma } \right)^{-1}.
\nonumber
\end{eqnarray}
For $Z\leq 3$, the fit parameters are given by: \\
 $ c_1= -0.349 + 1.766 \, Z  - 0.413 \, Z^2$,\\
 $ c_2= 3.499  -2.355  \, Z  + 0.871 \, Z^2$,\\
 $ c_3= 11.92  - 6.77  \, Z  + 1.04  \, Z^2$,\\
 $ c_4= 0.5733 -0.1116 \, Z  + 0.0159\, Z^2$,\\
 $ c_5= 0.7060 -0.9896 \, Z  + 0.3517\, Z^2$,\\
 $ c_6= 4.779  -2.676  \, Z  + 0.361 \, Z^2$,\\
 $ c_7= 2.457  +2.523  \, Z  -1.014  \, Z^2$.

For $Z>3$, we obtain
 $c_1= 1.232$, $c_2= 4.273$, $c_3= 0.9742$, $c_4= 0.3816$
 $c_5= 0.9025$, $c_6= 0.0$, $c_7= 0.9$.

If $Z=1$, the rms fit error over all our data set is about
$\delta_{\rm av}\approx $ 3.2\%, and the maximum
error of
$\delta_{\rm max} \approx$ 7.2\%
takes place at $\rho \approx 10^5$ g cm$^{-3}$ and $\Gamma=3$.
If $Z=2$ we have $\delta_{\rm av}\approx $ 3.7\%,
$\delta_{\rm max} \approx$ 7.8\%
at $\rho \approx 10^9$ g cm$^{-3}$ and $\Gamma=1$.
For $Z=3$, we obtain
$\delta_{\rm av} \approx $ 2.4\%,
$\delta_{\rm max} \approx$ 7.0\%
at $\log(\rho[{\rm g \; cm}^{-3}]) \approx 2.5$ and $\Gamma=1$.
For higher $Z$, the
rms error remains about 1.5--1.7\%,
and the maximum error mainly decreases. For instance,
we have $\delta_{\rm max} \approx$ 5.1\%
at $\log\rho \approx 2.25$ and $\Gamma=4$
for $Z=6$;
$\delta_{\rm max} \approx$ 3.7\%
at $\log\rho \approx 5.75$ and $\Gamma=2$
for $Z=12$; and
$\delta_{\rm max} \approx$ 3.4\%
at $\log\rho \approx 5.5$ and $\Gamma=8$
for $Z= 26$. This fit accuracy is quite sufficient
for studying the thermal structure of NSs.

%%%%%%%%%%%%%%%%%%%%%%%%%%%%%%%%%%%%%%%%%%%%%%%%%%%%%   APPENDIX C
\subsection{Relation between internal and effective temperatures}
\label{app-link}
In this section, we derive a
fitting formula for $T_{\rm eff}$ as a function of $T_{\rm b}$, 
valid for $4.7 \leq \log T_{\rm eff} \leq 6.5$,
$0.4 \leq g_{14} \leq 6$,
where $g_{14}$ is the surface gravity in units of $10^{14}$ cm s$^{-2}$.

Let us define $T_{\rm b9}=T_{\rm b}/10^9$~K, 
$T_{\rm eff6}=T_{\rm eff}/10^6$~K, and
\begin{equation}
   \eta\equiv g_{14}^2\Delta M/M,
\end{equation}
where
$M$ is the NS mass and $\Delta M$ is the accreted mass.
According to GPE,
$
   \eta=P_{\rm b}/(1.193\times 10^{22}\mbox{ Mbar}),
$
where $P_{\rm b}$ is the pressure at the bottom of the accreted envelope.

For a purely iron (non-accreted) envelope,
a very crude estimate (with an error $\sim$ 30\%) yields
\begin{equation}
   T_{\rm eff6}=T_{\ast}\equiv
   \left(7\,T_{\rm b9}\sqrt{g_{14}}\right)^{1/2}.
\end{equation}
Then  
$
   \zeta\equiv T_{\rm b9}-(T_\ast /10^3)
$
is approximately
an increase of the temperature through the iron envelope (in $10^9$~K).
A more accurate fitting formula reads
\begin{equation}
   T_{\rm eff6,Fe}^4=g_{14} \, [(7\zeta)^{2.25}+(\zeta/3)^{1.25}].
\label{t0}
\end{equation}
The typical fit error of
$T_{\rm eff6,Fe}$
is about 2\%, with maximum 4.2\%, 
over the $T_{\rm eff}-g$ domain indicated above. 

For a fully accreted envelope, we obtain
\begin{equation}
   T_{\rm eff6,a}^4=g_{14} \, (18.1\,T_{\rm b9})^{2.42},
\label{t1}
\end{equation}
which is valid at
not too high temperature ($T_{\rm b}\leq 10^8$~K).

Finally, for the partially
accreted envelopes at any temperatures
within the indicated range, we have
\begin{equation}
   T_{\rm eff6}^4={a\,T_{\rm eff6,Fe}^4+T_{\rm eff6,a}^4
   \over a+1},
\label{ttfit}
\end{equation}
where
\begin{equation}
   a=\left[1.2+(5.3\times 10^{-6}/\eta)^{0.38}\right]\,T_{\rm b9}^{5/3}.
\end{equation}
The typical fit
error of Eq.\,(\ref{ttfit}) for $T_{\rm eff6}$ 
is about 3\%, with maximum 5.2\%, 
for all possible values of $\eta$ and any values 
of $g$ and $T_{\rm eff}$ within the indicated range.

The dependence (\ref{t0}) is recovered
not only at sufficiently low accreted mass
($\eta\to0$), but also at sufficiently high $T_{\rm b}$. 
The latter result reflects 
the fact (first demonstrated by GPE) that at high $T_{\rm eff}$
the thermal insulation is mostly produced by the
conductive opacities in the deep and hot layers of the envelope,
in which light elements (H, He) burn into heavier ones.
On the other hand, even at very low accreted mass  
($\log\Delta M/M\sim -16$), 
the approximation of fully accreted crust is good enough 
at sufficiently low temperature
because in this case the thermal insulation is provided mainly by the
low-density accreted surface layers.


\begin{thebibliography}{}

\bibitem{}
Alcock C., Illarionov A.F., 1980, ApJ 235, 534

\bibitem{}
Anderson S.B., C\'{o}rdova F.A., Pavlov G.G., Robinson C.R., 
Thompson R.J., 1993, ApJ 414, 867

\bibitem{}
Ayasli S., Joss P.C., 1982, ApJ 256, 637

\bibitem{}
Baiko D.A., Yakovlev D.G., 1995, Astron.\ Lett. 21, 702.

\bibitem{}
Becker W.,  Aschenbach B., 1995, {\it ROSAT\/} HRI Observations
of the Crab Pulsar.
In: Alpar M.A., Kizilo\u{g}lu \"{U}.,
Paradijs J.\ van (eds.) The Lives of the Neutron Stars.
Kluwer, Dordrecht, p.\ 47

\bibitem{}
Bignami G.F., Caraveo P.A., Mignani R., Edelstein J., 
Bowyer S., 1996, ApJ 456, L111

\bibitem{}
Blaes O.M., Blandford R.D., Madau P., Yan L., 1992,
ApJ 399, 634

\bibitem{}
Boerker D.B., Rogers F.J., DeWitt H.E., 1982, Phys.\ Rev.\ A 
25, 1623

\bibitem{}
Brown G.E., Weingartner J.C., 1994, ApJ 436, 843

\bibitem{}
Chabrier G., 1990, Journal de Physique 51, 57

\bibitem{}
Chabrier G., 1993, ApJ 414, 695

\bibitem{}
Chabrier G., Schatzman E. (eds), 1994,
Proc.\ IAU Coll. 147, 
The Equation of State in Astrophysics.
Cambridge Univ.\ Press, Cambridge

\bibitem{}
Chevalier R.A., 1989, ApJ 346, 847

\bibitem{}
Chevalier R.A., 1996, ApJ 459, 322

\bibitem{}
DeWitt H.E., Slattery W.L., Chabrier G., 1996,
Physica B, 228, 21

\bibitem{}
Doggett J.A., Spencer L.V., 1956, Phys.\ Rev. 103, 1597

\bibitem{}
Ebeling W., Kraeft W.D.,  Kremp D., 1977, 
Theory of Bound States
and Ionization Equilibrium of Plasmas and Solids.
Akademie, Berlin

\bibitem{}
Ergma E., 1986, Sov.\ Sci.\ Rev.\ E:
Astrophys.\ Space Phys. 5, 181

\bibitem{}
Fletcher C.A.J., 1988, Computational Techniques for Fluid 
Dynamics. Springer, Berlin

\bibitem{}
Flowers E.,  Itoh N., 1976, ApJ 206, 218

\bibitem{}
Fontaine G., Graboske H.C., Van Horn H.M., 1977,
ApJS 35, 293 

\bibitem{}
Friman B.L.,  Maxwell O.V., 1979, ApJ 232, 541

\bibitem{}
Gnedin O.Y.,  Yakovlev D.G., 1993, Astron.\ Lett. 19, 104.

\bibitem{}
Greiveldinger C., Camerini U., Fry W. et al., 1996, 
ApJ 465, L35

\bibitem{}
Gudmundsson E.H., Pethick C.J.,  Epstein R.I., 1983, 
ApJ 272, 286 (GPE)

\bibitem{}
Hailey C.J., Craig W.W., 1995, ApJ 455, L151

\bibitem{}
Halpern J.P.,  Ruderman M., 1993, ApJ 415, 286

\bibitem{}
Hansen J.-P.,  Vieillefosse P., 1975, Phys.\ Lett. A 53, 187

\bibitem{}
Hernquist L., 1984, ApJS 56, 325

\bibitem{}
Hernquist L., 1985, MNRAS 213, 313

\bibitem{}
Hernquist L., Applegate J.H., 1984, ApJ 287, 244

\bibitem{}
Huebner W.F., Merts A.L., Magee N.H., Argo M.F., 1977,
Astrophysical Opacity Library. Rept.\ LA\,6760M, 
Los Alamos Scientific Laboratory (LAO)

\bibitem{}
Hubbard W.B.,  Lampe M., 1969, ApJS 18, 297

\bibitem{}
Iben I., 1974, ARA\&A 12, 215

\bibitem{}
Itoh N., Mitake S., Iyetomi H.,  Ichimaru S., 1983, ApJ 273, 774

\bibitem{}
Itoh N., Kohyama Y., Matsumoto N.,  Seki M., 1984, ApJ 
   285, 758; erratum 404, 418

\bibitem{}
Jancovici B., 1962, Nuovo Cimento 25, 428

\bibitem{}
Kippenhahn R., Weigert A., 1990, Stellar Structure and Evolution.
Springer, Berlin

\bibitem{}
Lampe M., 1968, Phys.\ Rev. 170, 306

\bibitem{}
Landau L.D.,  Lifshitz E.M., 1986,
Statistical Physics, Part I.
Pergamon, Oxford

\bibitem{}
Lattimer J.M., Pethick C.J., Prakash M.,  Haensel P., 1991,
Phys.\ Rev.\ Lett. 66, 2701

\bibitem{}
Levenfish K.P.,  Yakovlev D.G., 1996, Astron.\ Lett. 22, 56

\bibitem{}
Lyne A.G., Pritchard R.S., Graham-Smith F., Camilo F., 1996, 
Nature 381, 497

\bibitem{}
Meyer R., Pavlov G.G., M\'{e}sz\'{a}ros P., 1994, ApJ 433, 265

\bibitem{}
Miralda-Escud\'{e} J., Haensel P., 
Paczy\'{n}ski B., 1990, ApJ 362, 572

\bibitem{}
Morley P.D., 1996, A\&A 313, 204

\bibitem{}
Muslimov A., Page D., 1995, ApJ 440, L77

\bibitem{}
Nagara H., Nagata Y.,  Nakamura T., 1987, 
Phys.\ Rev. A36, 1859

\bibitem{}
Nelson R.W., Salpeter E.E., Wasserman I., 1993, ApJ 418, 874

\bibitem{}
Nomoto K.,  Tsuruta S., 1987, ApJ 312, 711

\bibitem{}
\"{O}gelman H., Finley J.P., Zimmermann H., 1993, 
Nature 361, 136

\bibitem{}
\"{O}gelman H., 1995, X-ray Observations of Cooling Neutron 
Stars.
In: Alpar M.A., Kizilo\u{g}lu \"{U}.,
Paradijs J.\ van (eds.) The Lives of the Neutron Stars.
Kluwer, Dordrecht, p.\ 101

\bibitem{}
Paczy\'{n}ski B., 1983, ApJ 264, 282

\bibitem{}
Page D., 1994, ApJ 428, 250

\bibitem{}
Page D., 1996, ApJ Lett. (submitted)

\bibitem{}
Page D., Shibanov Yu.A., Zavlin V.E. 1996,
Temperature, Distance and Cooling of the Vela Pulsar.
In: Zimmermann H.-U., Tr\"umper J., Yorke H. (eds) 
Proc.\ Int.\ Conf.\ on X-ray Astronomy and Astrophysics, 
R\"ontgenstrahlung from the Universe.
MPE Report 263, Garching, p.\ 173
 
\bibitem{}
Pavlov G.G., Potekhin A.Y., 1995, ApJ 450, 883

\bibitem{}
Potekhin A.Y., 1996, Physics of Plasmas (in press)

\bibitem{}
Potekhin A.Y., Yakovlev D.G., 1996, A\&A 314, 341 (PY).

\bibitem{}
Prakash M., Ainsworth T.L.,  Lattimer J.M., 1988, 
Phys.\ Rev.\ Lett. 61, 2518

\bibitem{}
Rajagopal M., Romani R.W., 1996, ApJ 461, 327

\bibitem{}
Rogers F.J., Swenson F.J.,  Iglesias C.A., 1996, 
ApJ 456, 902

\bibitem{}
Saumon D.,  Chabrier G.,  Van Horn H.M., 1995,
ApJS 99, 713 (SCVH)

\bibitem{}
Schaaf M.E., 1988, A\&A 205, 335

\bibitem{}
Schaaf M.E., 1990, A\&A 235, 499

\bibitem{}
Schramm S., Langanke K.,  Koonin S.E., 1992, ApJ 397, 579

\bibitem{}
Schwarzschild M., 1958, Structure and Evolution of the Stars. 
Princeton Univ.\ Press, Princeton

\bibitem{}
Segretain L., Chabrier G., Hernanz M., et al., 1994,
ApJ 434, 641

\bibitem{}
Shibanov Yu.A.,  Yakovlev D.G., 1996, A\&A 309, 171

\bibitem{}
Stringfellow G.S., DeWitt H.E.,  Slattery W.L., 1990,
Phys.\ Rev.\ A 41, 1105

\bibitem{}
Thorne K.S., 1977, ApJ 212, 825

\bibitem{}
Timmes F.X., 1992, ApJ 390, L107

\bibitem{}
Umeda H., Tsuruta S.,  Nomoto K., 1994, ApJ 433, 256

\bibitem{}
Urpin V.A.,  Yakovlev D.G., 1980, SvA 24, 126

\bibitem{}
Van Riper K.A., 1988, ApJ 329, 339

\bibitem{}
Van Riper K.A., 1991, ApJS 75, 449

\bibitem{}
Williams R.H., DeWitt H.E., 1969, Phys.\ Fluids 12, 2326

\bibitem{}
Yakovlev D.G., 1987, 
SvA 31, 347

\bibitem{}
Yakovlev D.G., 1994, Acta Phys.\ Polonica 25B, 401

\bibitem{}
Yakovlev D.G.,  Kaminker A.D., 1994, 
Neutron Star Crusts with Magnetic Fields. 
In: Chabrier G., Schatzman E. (eds) 
Proc.\ IAU Coll. 147, 
The Equation of State in Astrophysics.
Cambridge Univ.\ Press, Cambridge,
p.\ 214

\bibitem{}
Yakovlev D.G.,  Levenfish K.P., 1995, A\&A 297, 717

\bibitem{}
Yakovlev D.G.,  Shalybkov D.A., 1989, Sov.\ Sci.\ Rev.\ E:
Astrophys.\ Space Phys. 7, 313

\bibitem{}
Yakovlev D.G.,  Urpin V.A., 1980, SvA 24, 303 (YU)

\bibitem{}
Zavlin V.E., Pavlov G.G., Shibanov Yu.A., Rogers F.J.,
Iglesias C.A., 1996, 
X-ray Spectra from Convective Photospheres of Neutron Stars. 
In: Zimmermann H.-U., Tr\"umper J., Yorke H. (eds) 
Proc.\ Int.\ Conf.\ on X-ray Astronomy and Astrophysics, 
R\"ontgenstrahlung from the Universe.
MPE Report 263, Garching, p.\ 209

\end{thebibliography}
\end{document}